\documentclass[12pt,preprint]{aastex}
\shorttitle{Ultra-luminous infrared galaxies in SDSS DR6}
\shortauthors{ Hou, Wu \& Han}
\begin{document}
\newcommand{\vdag}{(v)^\dagger}
\newcommand{\myemail}{skywalker@***}

   \title{Ultra-luminous Infrared galaxies in Sloan Digital
Sky Survey Data Release 6}

\author{L.~G. Hou$^1$, Xue-Bing Wu$^2$, J.~L. Han$^1$ }

 \altaffiltext{1}{National Astronomical Observatories, Chinese Academy
   of Sciences, Jia-20, DaTun Road, Chaoyang District, Beijing 100012,
   China}
\altaffiltext{2}{Department of Astronomy, School of Physics,
   Peking Universty, Beijing 100871, China}

\begin{abstract}
%
  Ultra-luminous infrared galaxies (ULIRGs) are interesting objects
  with dramatic properties. Many efforts have been made to understand
  the physics of their luminous infrared emission and evolutionary
  stages. However, a large ULIRG sample is {still} needed to
  study the properties of their central black holes (BHs), the
  BH$-$host galaxy relation, and their evolution. We identified
  {308 ULIRGs from the Sloan Digital Sky Survey Data Release
    6}, and classified them into {the} NL ULIRGs (with only
  narrow emission lines) and {the} Type I ULIRGs (with broad
  emission lines). About {56\%} of ULIRGs in our total sample
  show interaction features, and this percentage is {79\%} for
  redshift $z$ $<$ 0.2. Optical identifications of these ULIRGs show
  that the active galactic nucleus percentage is at least {49\%}, and the
  percentage increases with {the} infrared luminosity. We found
  {62} Type I ULIRGs, and estimated their BH masses and
  velocity dispersions from their optical spectra. Together with
  {known} Type I ULIRGs in the literature, {a sample of
    90 Type I ULIRGs enables us to make a statistical study}. We found
  that the BH masses of Type I ULIRGs are typically smaller than those
  of Palomar$-$Green (PG QSOs), and most Type I ULIRGs follow the $M_{\rm BH}$$-$$\sigma$ relation. However, some ULIRGs with a larger
  Eddington ratio {deviate from this} relation, even though the line
  width of the [O$_{\rm III}$] narrow-line (NL) core or the [S$_{\rm II}$] line was used as the
  surrogate of velocity dispersion. This implies that at least some
  ULIRGs are probably still in the early {evolution} stage
  toward QSOs. {The} anti-correlation {between} the mass
  deviation from the $M_{\rm BH}$$-$$\sigma$ relation {and} the
  Eddington ratio supports that the evolution of Type I ULIRGs
  {is} probably followed by the building up of the $M_{\rm BH}$$-$$\sigma$ relation and the evolution to the QSO phase.

\end{abstract}
%
{}

\keywords{galaxies: active --- galaxies: formation --- galaxies:
nuclei
--- galaxies: starburst --- infrared: galaxies}

%

\section{Introduction}

\label{sect:intro} Ultra-luminous infrared galaxies (ULIRGs) were
discovered by \emph{Infrared Astronomical Satellite (IRAS)} in large
numbers with infrared luminosity in 8$-$1000 ${\rm {\rm \mu m}}$
greater than 10$^{12}$ $L_{\odot}$. The {serious intrinsic}
obscuration {for
  ULIRGs} in optical, UV, and even in mid-IR and X-ray bands
\citep{chyt91} makes it difficult to clearly probe their physical
properties by observations. There are probably more ULIRGs at high
redshifts than {in the local universe}, and even more than optically
bright QSOs \citep{lfs06}. In the last two decades, multi-wavelength
studies on ULIRGs have significantly improved our understanding of
these dramatic objects \citep[see the review papers of][]{sm96,lfs06}.

Extremely {high} infrared luminosity of ULIRGs is dominated by
starbursts, and sometimes with additional contribution from active
galactic nuclei (AGNs). Most ULIRGs are interacting systems
undergoing a wide range of merger stages
\citep[e.g.,][]{zxds91,csm+96,mam+96,sse00,fro+01,vks02}. ULIRGs
with large luminosity or spectroscopic signatures of AGNs are most
likely late-stage mergers \citep{vks02}. Observations of the
molecular gas in ULIRGs \citep[see][]{ms88,ssy+86,sss91,gs04} proved
that high-density gases are reserved in a compact nuclear region.
The connection between ULIRGs and AGNs was also found from optical
and mid-infrared spectra \citep[e.g.][]{kvs98,vsk97,vks99,vsk99}.
About 25$\%$ of ULIRGs present evidences of AGNs, and the percentage
increases to $\sim$ 50\% when $L_{\rm IR}$ is greater than
10$^{12.3}$ $L_{\odot}$. Less than 10\% ULIRGs have broad emission
lines, which are the so-called Type I ULIRGs
\citep[see][]{csm+96,wzxd981,wzxd98,zwm+99,cs01,cxd+01}. ULIRGs and
AGNs  most probably have evolutionary connection. To understand such
possible connection, \citet{lfs06} suggest that major mergers of
gas-rich galaxies first form a massive cool starburst-dominated
ULIRG, and then a warm ULIRG phase is followed when a central AGN
turns on inside the dust cocoon and heats the surrounding dust. The
central AGN will evolve into an optically bright phase when it blows
away the surrounding dust cocoon. The resulting stellar system will
resemble a spheroid, so that the mass of the central black hole (BH)
of the AGN ($M_{\rm BH}$) is related to the stellar velocity
dispersion $\sigma$. The $M_{\rm BH}$$-$$\sigma$ relation might
build up at that time. In this scenario, ULIRGs are in a pre-AGN
phase. The typical BH mass of ULIRGs should be smaller, and the
galaxy bulge may not have been constructed completely, compared with
those of normal QSOs.

\citet{kan+06} used a sample of eight Type I ULIRGs in the local
universe with data of {full width at half-maximum (FWHM)} of H$_{\rm
  \beta}$ and {optical continuum luminosity} $L_{\rm
  \lambda}({\rm 5100{\AA}})$ from \citet{zxm+02} to estimate their BH
masses and investigate the BH$-$bulge relation. They found that the
Type I ULIRGs have systematically smaller BH masses in spite of
having the comparable bulge luminosity as QSOs and elliptical
galaxies. We note that the FWHM of H$_{\rm \beta}$ given in
\citet{zxm+02} is the FWHM of the whole emission-line profile, not
the FWHM of the broad-line component, which should be used in the
estimation of the BH mass. Therefore, \citet{kan+06} may
underestimate the masses of BHs for these ULIRGs. Using a sample of
sources mostly from \citet{zxm+02}, \citet{hxm+05} carried out a
study on the Type I ULIRGs (named as IR QSOs in their paper), and
concluded that the typical BH mass of Type I ULIRGs is smaller, and
the typical Eddington ratio ($L_{\rm bol}$/$L_{\rm Edd}$) is larger
than those of PG QSOs. At higher redshift, \citet{bsc+05} found that
submillimeter galaxies have smaller BH masses than QSOs with respect
to the same mass range of bulges. \citet{abs+98} concluded that
submillimeter galaxies host BHs with a mass of $\log({M_{\rm
BH}/M_{\odot}})\approx 7.8$ in their sample. Because local ULIRGs
and high-redshift submillimeter galaxies are similar with each
other, both having bright infrared luminosity and a large amount of
gas, and locating in the interacting systems, the investigations on
local ULIRGs could enlighten our understanding about the
high-redshift submillimeter galaxies.

A large ULIRG sample is needed to study their central BHs, the
BH$-$host galaxy relation, and galaxy evolution. The best-known
samples of \emph{IRAS} luminous infrared galaxies and ULIRGs are the
Bright Galaxy Sample of \citet{ssm+87}, updated into the Revised
Bright Galaxy Sample by \citet{smk+03}, the complete flux-limited
\emph{IRAS} 1 Jy sample \citep{ks98}, the 2 Jy sample of
\citet{sdyh90}, and the FIRST/\emph{IRAS} sample of \citet{ssbb00}.
Since the Sloan Digital Sky Survey (SDSS) covers more than a quarter
of the sky, more ULIRGs with {relatively} high quality spectra can
be found by the cross-correlation of \emph{IRAS} data with the SDSS.
\citet{goto05} investigated the optical properties of 4248 infrared
galaxies with $L_{\rm IR}$ of $10^9$ to 10$^{13.57}$ $L_{\odot}$
from the cross-correlation between SDSS DR3 spectroscopic sample of
galaxies and \emph{IRAS} sources, and 181 of them are ULIRGs.
\citet{pkh05} used the SDSS DR2 data to study the optical properties
of \emph{IRAS} galaxies. \citet{cww+06} identified 1207 luminous
infrared galaxies and 57 ULIRGs from SDSS DR2 for a statistical
study. \citet{hsl+07} identified 324 ULIRGs from the SDSS, 2dF
Galaxy Redshift Survey \citep{cdm+01}, and 6dF Galaxy Survey
\citep{jsc+04}. Because the new SDSS spectroscopic sample of
galaxies in DR6 has been released \citep{adel08}, a larger ULIRG
sample can be identified and then used to re-examine the statistical
properties of ULIRGs.

This paper is organized as follows. In Section 2, we make a
cross-identification between the \emph{IRAS} Faint Source Catalog
and the spectroscopic catalog of the SDSS DR6, and obtain a ULIRG
sample. We separate them into two sub-samples, NL ULIRGs and Type I
ULIRGs, and fit their SDSS spectra. We present the results and
discuss the NL ULIRG sample in Section 3. In Section 4, we carry out
a study on the BH masses and the $M_{\rm BH}$$-$$\sigma$ relation of
Type I ULIRGs. The discussions and conclusions are presented in
Sections 5 and 6. In this paper, we adopt $H_0$ = 70 km s$^{-1}$
Mpc$^{-1}$, $\Omega_{\rm m}$ = 0.3, and $\Omega_{\rm \Lambda}$ =
0.7.


\section{The ULIRG sample}
\label{sect:Obs}

SDSS DR6 spectra catalog contains {about 750,000 galaxies and QSOs}
and covers over 7425 deg$^2$ \citep{adel08}. \emph{IRAS} Faint
Sources Catalog \citep[][hereafter FSC92]{mkc92} contains 173,044
sources with IR flux in bands of 12, 25, 60, and 100 ${\rm \mu m}$.
Although the \emph{IRAS} data set was published more than 17 years
ago, it is still very underexplored. Only 43\% of the total
\emph{IRAS} extragalactic FSC sources have been included in any sort
of publication \citep[][]{lfs06}.

The positional uncertainty of \emph{IRAS} source (about 1$-$13
arcsec for the in-scan direction and 3$-$55 arcsec for the
cross-scan direction) is much larger than that of the object in the
SDSS DR6, and is described by an uncertainty ellipse. {Similar to
\citet{hsl+07} and
  \citet{cww+06}}, we used the positional uncertainty ellipse of each
\emph{IRAS} source to obtain their matched counterparts in the SDSS.
If a galaxy of the SDSS DR6 falls into the 3$\sigma$ uncertainty
ellipse of \emph{IRAS} source, we regarded them as a match. As a
result, we found that {11,354} \emph{IRAS} sources have only one
counterpart in the SDSS DR6 and {984} have more than one
counterparts. In the later case, the likelihood ratio method
\citep[see][]{ss92,hsl+07,cww+06} is used to determine which
counterpart of this \emph{IRAS} source is the most probable one.
After doing these, we obtained a list of {12,338} \emph{IRAS}
sources with SDSS DR6 optical counterparts.

\subsection{Selection criteria}

Our selection criteria of ULIRGs are shown as follows: for the
{12,338} matched \emph{IRAS} sources, {their} 12 ${\rm \mu m}$ and
25 ${\rm \mu m}$ flux densities {are mostly} upper limits.
Therefore, we calculated their far-infrared luminosity by using {60
$\mu m$ and 100 $\mu m$ fluxes with the} following formulae
\citep[see][]{hkmb88,sm96} and converted it to the total infrared
luminosity \citep{cab+00}:
\begin{equation}
  {F_{\rm FIR}=1.26\times10^{-14}\{2.58f_{60}+f_{100}\}(\rm W~m^{-2})} ,
\end{equation}
\begin{equation}
 {L_{\rm FIR}=4\pi D_{\rm L}^2F_{\rm FIR}(L_{\odot})} ,
\end{equation}
\begin{equation}
 {L_{\rm IR}(1-1000 \mu m)=1.75L_{\rm FIR}} .
\end{equation}

{Here $f_{60}$, $f_{100}$ are the \emph{IRAS} flux densities in Jy
at
  60 and 100 $\mu m$, $D_{\rm L}$ is the luminosity distance, $F_{\rm
    FIR}$ is the far-infrared flux, $L_{\rm FIR}$ is the far-infrared
  luminosity, and $L_{\rm IR}$ is the infrared luminosity in
  $L_{\odot}$. For all the matched sources, we also required that the
  60 $\mu m$ flux with high quality measurement (In the FSC92, high quality,
  moderate quality, and upper limits of the flux measurements are
  marked as 3, 2, and 1, respectively). Because the 100 $\mu m$ flux
  does not affect much on the value of $L_{\rm IR}$
  \citep[see][]{cww+06}, we do not set the limit to the quality of the 100
  $\mu m$ flux density. }The redshift
confidence of each source should be larger than 0.65. {Finally} we
identified {325} ULIRG candidates with an IR luminosity greater than
10$^{12}$ $L_{\odot}$. When we check these {325} ULIRG candidates by
using the NASA/IPAC Extragalactic Database (NED), we found {18}
sources whose redshifts provided in the NED are not consistent with
our results. {According to the NED
  identifications, nine of these 18 sources have a Petrosian $r$-band
  magnitude less than 15, so they are too bright for the SDSS. If}
calculated their $L_{\rm IR}$ with the redshifts provided in the
NED, we found that {17 of 18 sources are not ULIRGs. This
discrepancy
  is probably due to the large position error of the \emph{IRAS} sources and
  the incompleteness of the SDSS spectra. The fraction of the sources
  which are not excluded in the examination with the NED (307/325) is
  about 94.5\%, which is consistent with the reliability of our sample
  (about 93.4\%, see the next section).  By excluding these 17 objects,
  we obtain a sample of 308 ULIRGs.} Detailed information of our
  ULIRGs is given in Table 1.

\begin{figure}
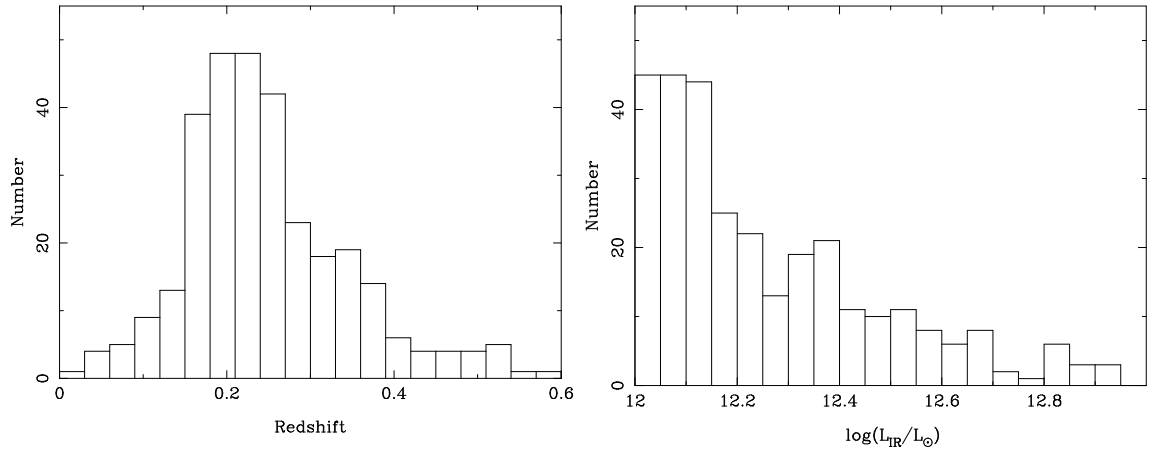

\centering \includegraphics[angle=270,width=75mm]{sdss_z.ps}
\centering \includegraphics[angle=270,width=75mm]{ULIRGs_LIR.ps}
\caption{Distribution of redshift (left panel) and $L_{\rm IR}$
  (right panel) of our ULIRG sample. }
\label{zlir}
\end{figure}

The distributions of redshift and infrared luminosity of ULIRGs are
shown in Figure~\ref{zlir}. The redshifts of our ULIRG sample cover
a range from about 0.03$-$0.6, with a median value of about 0.2,
which is similar to that of \citet{hsl+07}.

We compared our result with that of \citet{hsl+07}, who identified
126 ULIRGs from SDSS DR4 but adopted different methods to calculate
$L_{\rm
  IR}$. Our catalog recovers 122 of their 126 ULIRGs. For the
{rest of the four sources in their sample}, F07568+4823 and
F11553+4557 have IR luminosities of 10$^{11.99}$ $L_{\odot}$ in our
result and thus are not selected; {F10200+4839 and F15239+4331 are
not ULIRGs,
  and wrong optical counterparts were identified by \citet{hsl+07}.}

In order to obtain the radio properties of these ULIRGs, we
{also cross identified our ULIRG sample with the NVSS
  \citep{ccg+98} and FIRST \citep{bwh95, wbhg97} catalogs. Note that
  \citet{bkhi05} used a hybrid NVSS$-$FIRST method to identify the radio
  counterparts for SDSS DR2 galaxies with high reliability and
  completeness. We followed their method to identify radio
  counterparts of our ULIRG sample, and found that 140 of 308 ULIRGs
  have counterparts in the NVSS catalog within a typical searching
  radius about 15$^{\prime\prime}$, and 132 of these 140 sources have FIRST
  counterparts within 3$^{\prime\prime}$. Some of them are probably core-dominant
  radio sources, such as F08201+2801, F13408+4047, F16413+3954,
  F10418+1153, F09105+4108, F13451+1232, F11206+3639 and
  F08507+3636. For the rest of the 168 ULIRGs, 82 of them have one
  FIRST counterpart within 3$^{\prime\prime}$, and their radio flux densities are
  below or close to the NVSS flux limit (about 2.5 mJy). The radio
  information of these 222 ULIRGs is listed in our ULIRG catalog (see
  Table 1).}

\subsection{Reliability estimated with the likelihood ratio method}

To estimate the reliability of our sample, we follow
 \citet{cww+06} and \citet{hsl+07} and adopt a likelihood ratio
method \citep{ss92}. The likelihood ratio $p$ is defined as
\begin{equation}
  p{=\frac{Q(\leq m_r)\exp(-R^2\rm/2)}{2\pi{\sigma}_a{\sigma}_bn(\leq m_r)} }
  ,
\end{equation}
where $Q({\leq m_r})$ is the multiplicative factor {measuring} the
probability {for} a true optical counterpart brighter than the flux
limit exists in the association, and we set $Q$ = 1 for simplicity.
$\sigma_{\rm a}$ and $\sigma_{\rm b}$ are the standard deviations,
and ${\rm m_r}$ is the SDSS $r$-band magnitude. Here we assume that
the errors are Gaussian {distributed}(the error of an \emph{IRAS}
source is not a pure Gaussian, but in the statistical sense, the
result of likelihood ratio study can still be used to evaluate the
reliability), and define $R$ as
\begin{equation}
  {R^2=\frac{(d_1)^2}{{\sigma_{a1}}^2+{\sigma_{a2}}^2}+\frac{(d_2)^2}{{\sigma_{b1}}^2+{\sigma_{b2}}^2}} ,
\end{equation}
where $d_1$ and $d_2$ are the positional differences along the two
axes of the error ellipse between each \emph{IRAS} source and its
SDSS counterpart, $\sigma_{\rm a1}$ and $\sigma_{\rm b1}$ are the
errors of each \emph{IRAS} source along the $x$- and $y$-axes,
$\sigma_{\rm a2}$ and $\sigma_{\rm b2}$ are the errors of each
matched SDSS source along the $x$- and $y$-axes. Because the
positional error of an SDSS source is much smaller than that of an
\emph{IRAS} source, we {only consider
  $\sigma_{\rm a1}$ and $\sigma_{\rm b1}$ .}

\begin{figure}
\centering \includegraphics[angle=270,width=75mm]{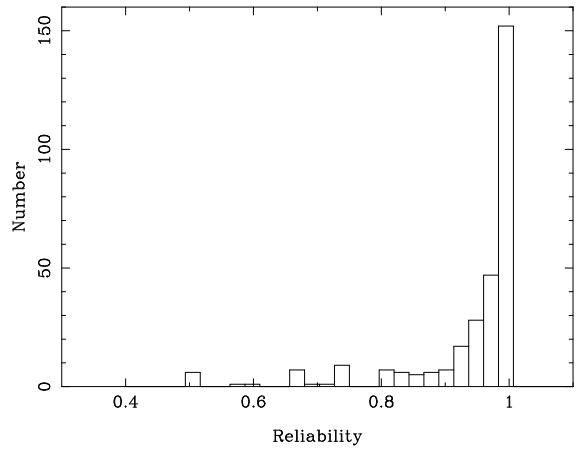}
\caption{Distribution of source matching reliabilities of our
  ULIRG sample.}
\label{reli}
\end{figure}

In this work, we adopt {the} 3$\sigma$ error ellipse as the match
justification; thus $n({\leq m_r})$ can be obtained by using the
formula:
\begin{equation}
 {n(\leq m_r)=\frac{N(\leq m_r)}{9\pi{\sigma}_{a1}{\sigma}_{b1}}} ,
\end{equation}
{where $n({\leq m_r})$ is the total surface density of objects
  brighter than the candidate, and $N({\leq m_r})$ is} the number of
galaxies whose magnitude is less than or equal to ${\rm m_r}$. Under
the above considerations, we obtain
\begin{equation}
   p{=\frac{9\exp(-R^2/2)}{2N(\leq m_r)}} ,
\end{equation}
where we use the $r$-band Petrosian magnitude for galaxies of the
SDSS DR6 to calculate $p$ of each source. To obtain the reliability,
we adopt the method proposed by \citet{lce+98} and \citet{mcb+01}.
The reliability of a source with $p$ is given by
\begin{equation}
   Re(p)=1-\frac{N_{random}(p)}{N_{true}(p)} ,
\end{equation}
where $N_{true}(p)$ represents the number of true associations, and
$N_{random}(p)$ represents the number of random associations with a
$p$, which can be derived by offsetting the positions of \emph{IRAS}
sources and re-calculating the associated sample. The numbers of
true and random-matched sources are {12,338} and {813}, and then the
reliability of our sample is about {93.4\%}. Thus we believe that
our ULIRG sample is reliable enough to make a statistical study on
the properties of NL ULIRGs and Type I ULIRGs. The distribution of
the reliabilities for our {308} ULIRGs is shown in
Figure~\ref{reli}.

\begin{figure*}
\centering \includegraphics[width=40mm]{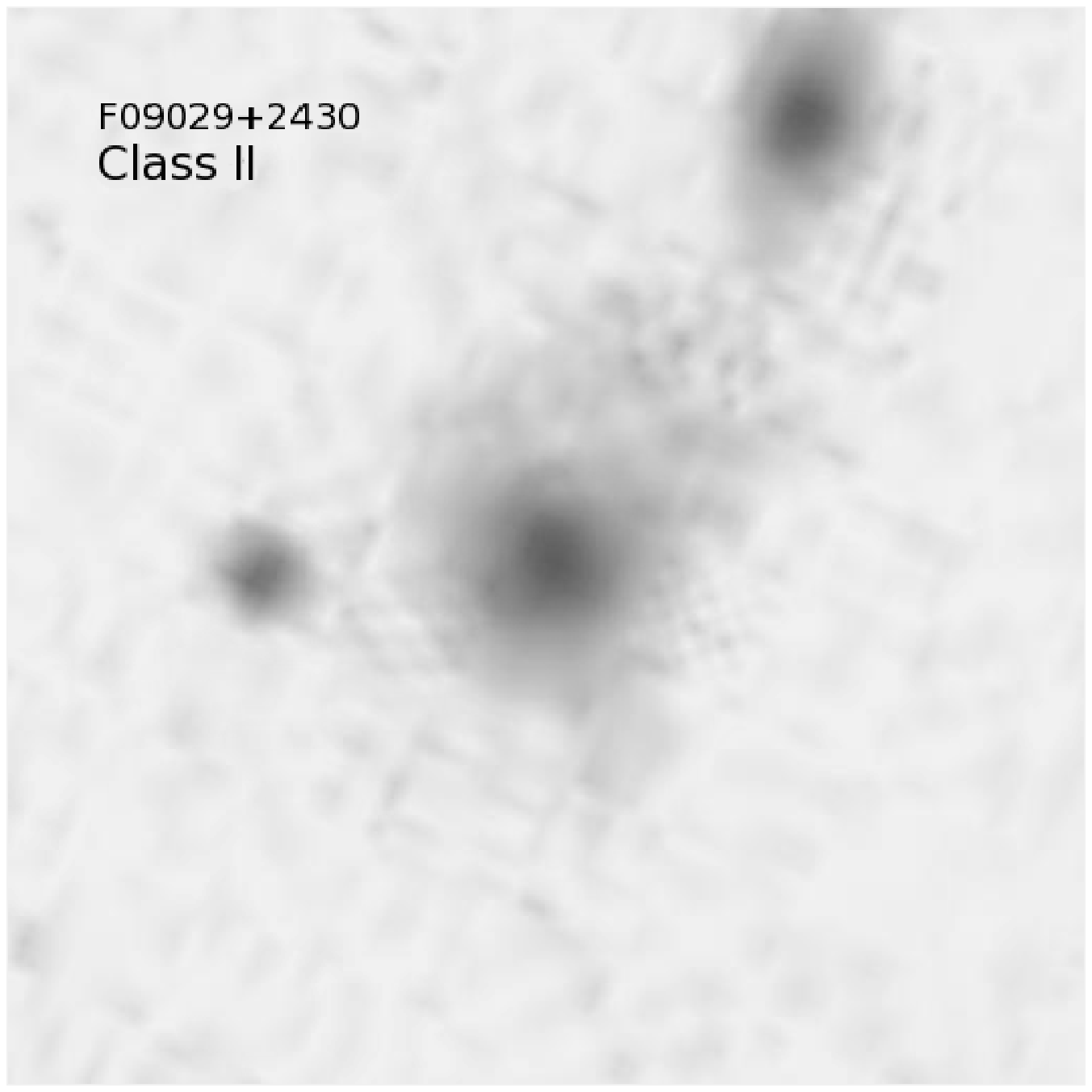}
\centering \includegraphics[width=40mm]{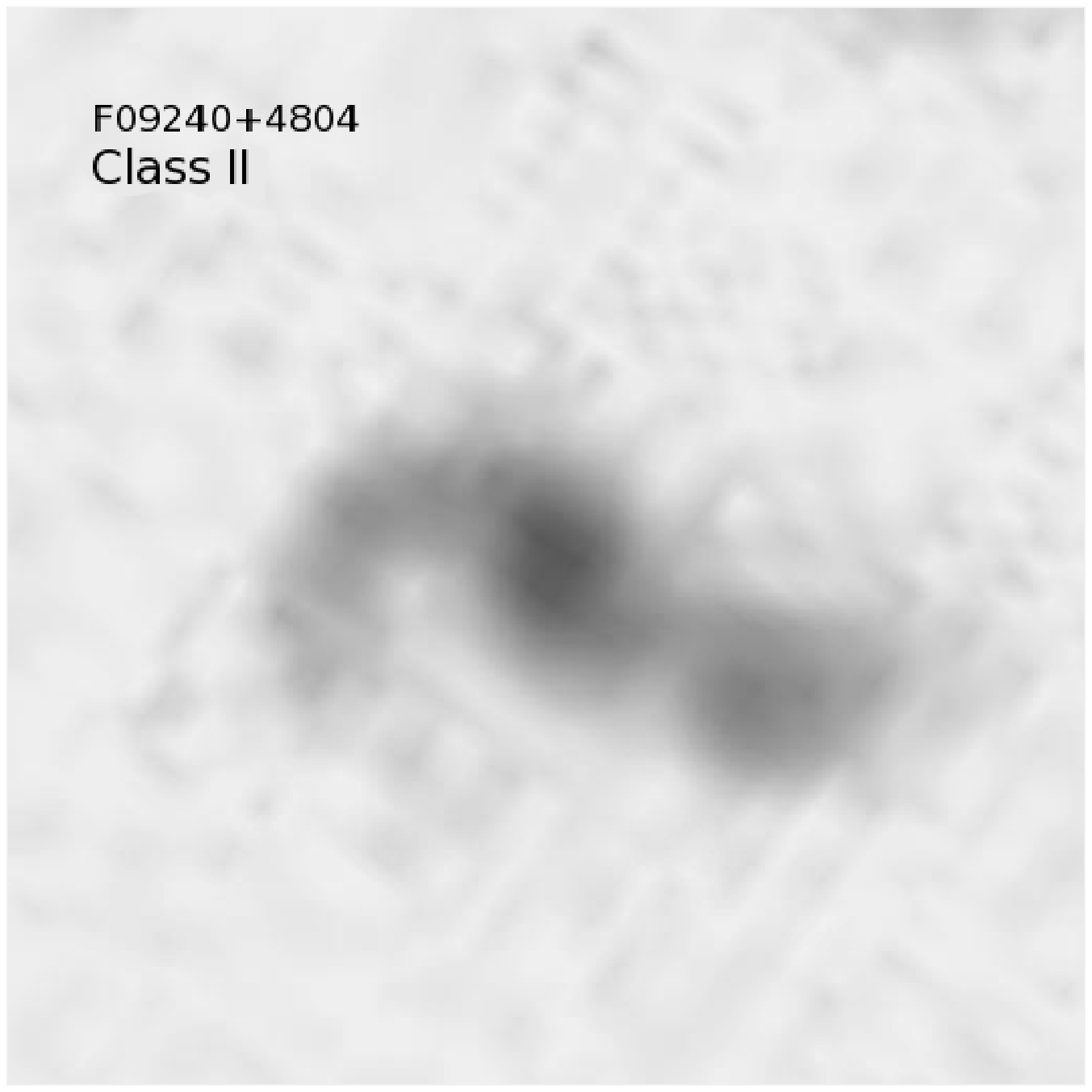}
\centering \includegraphics[width=40mm]{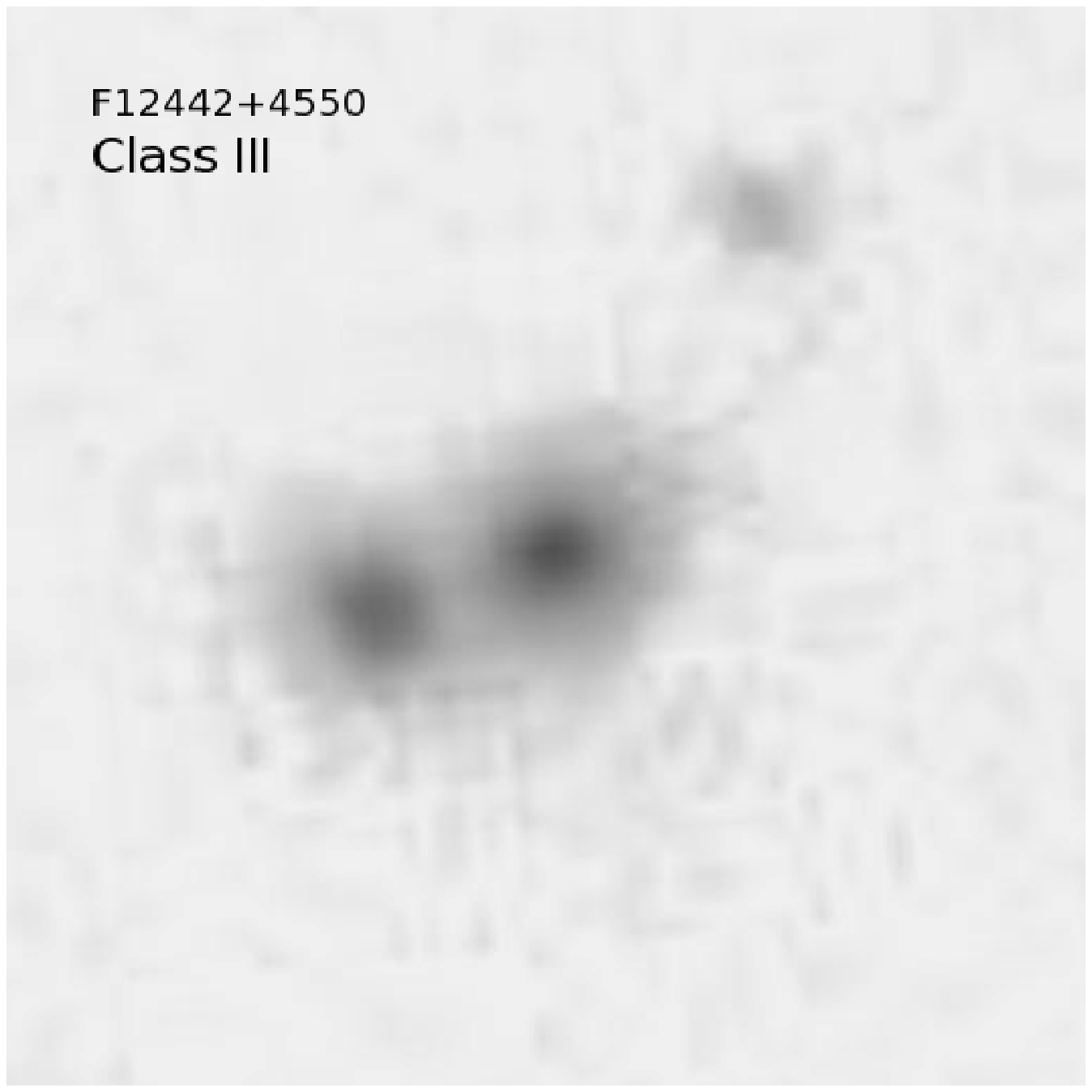}
\centering \includegraphics[width=40mm]{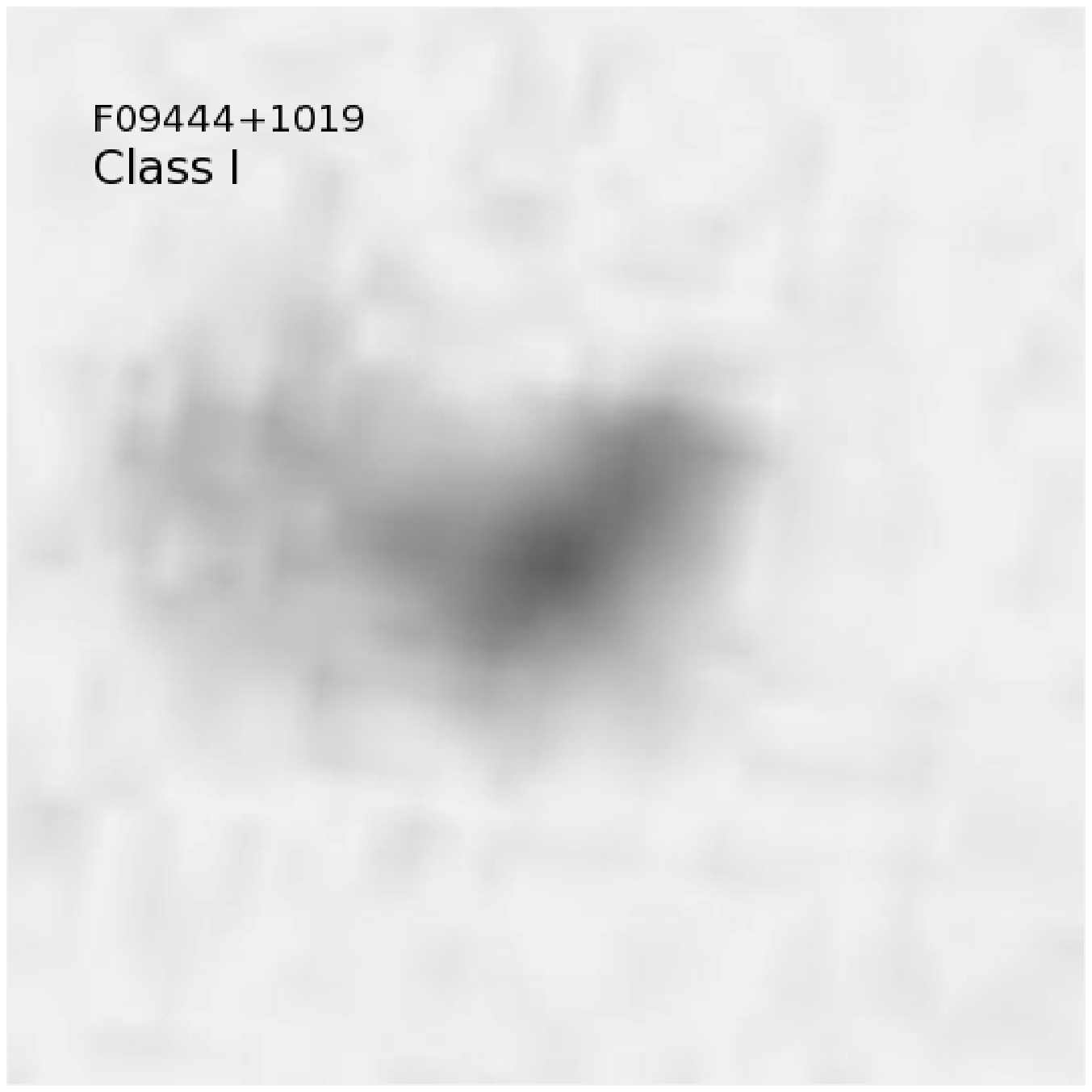}\\
\centering \includegraphics[width=40mm]{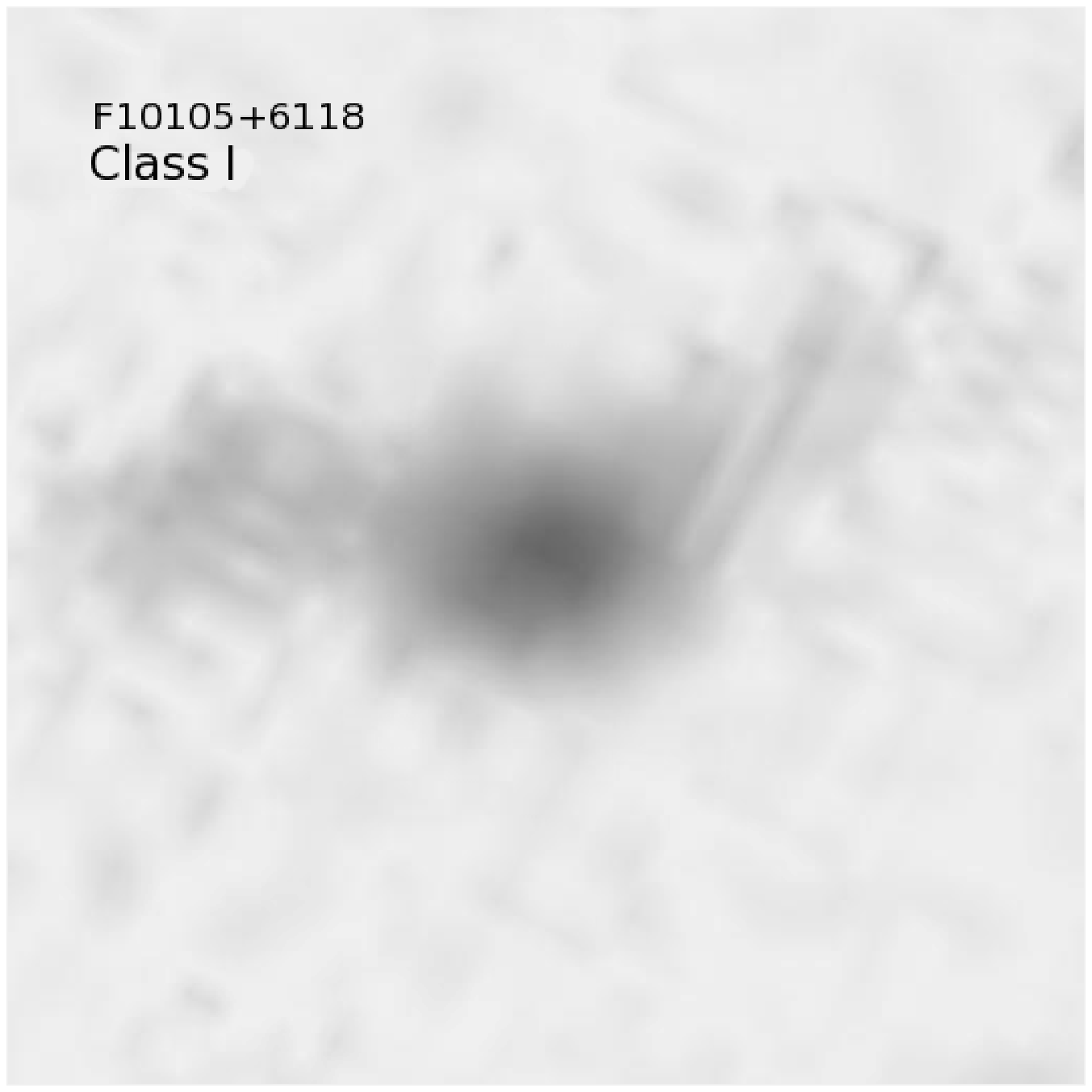}
\centering \includegraphics[width=40mm]{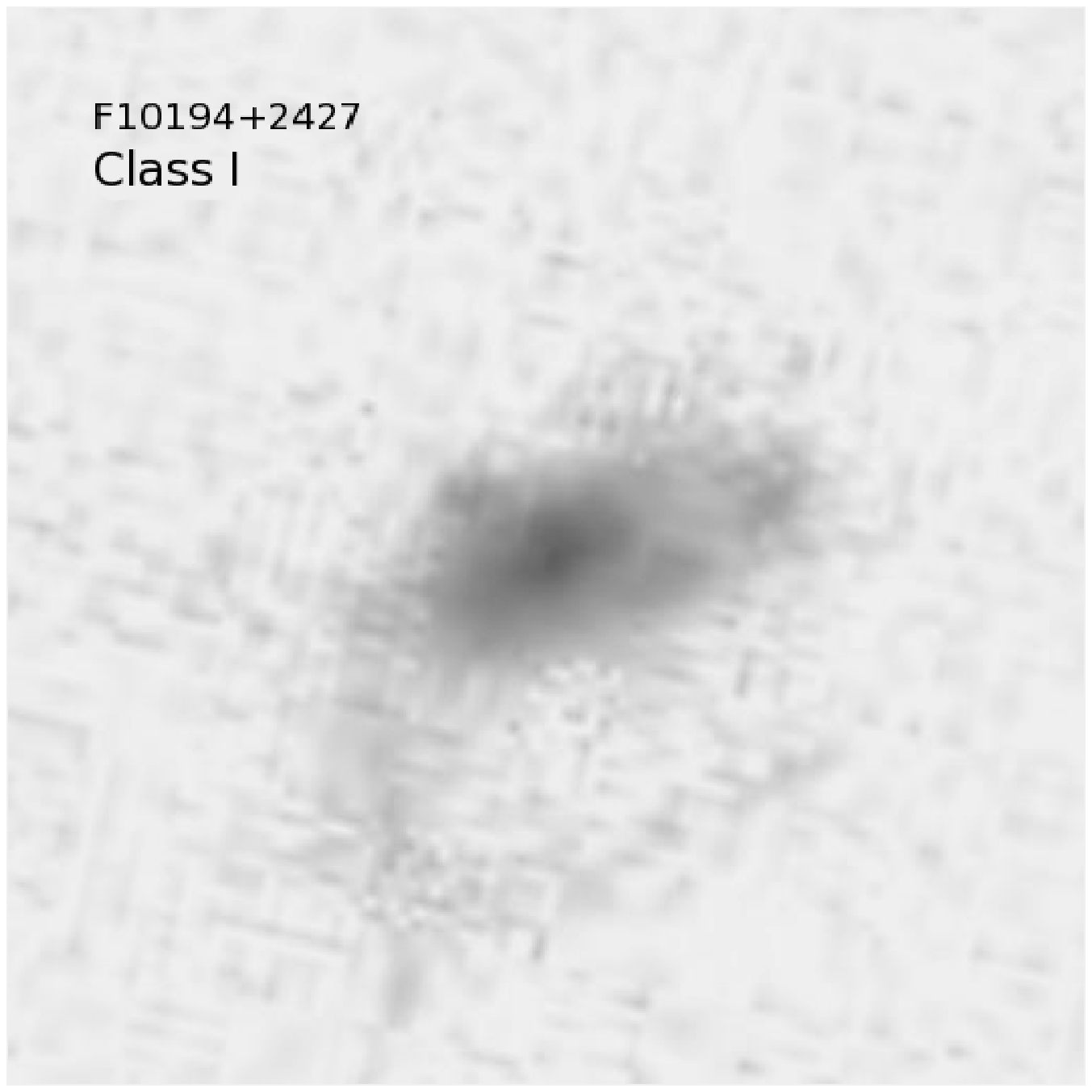}
\centering \includegraphics[width=40mm]{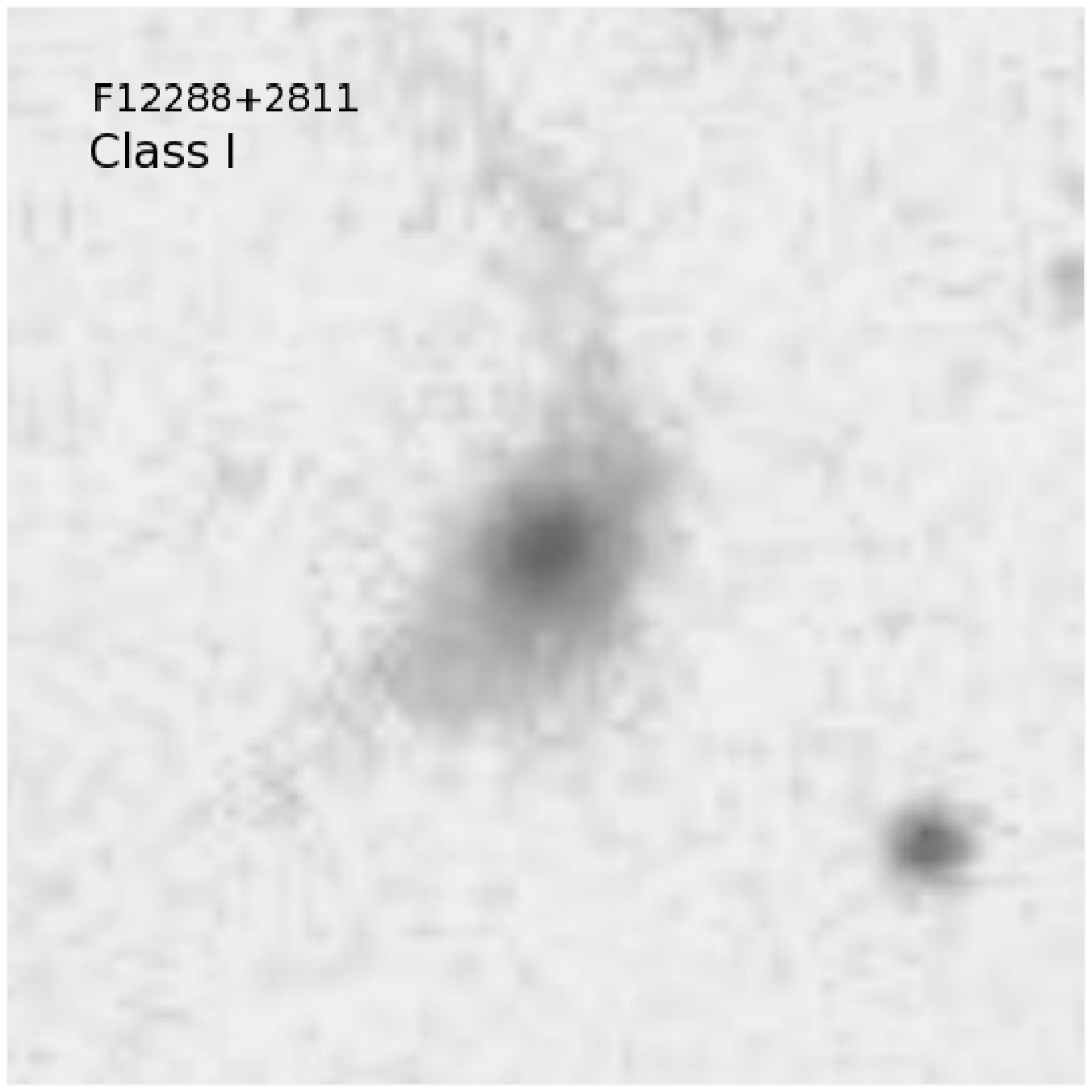}
\centering \includegraphics[width=40mm]{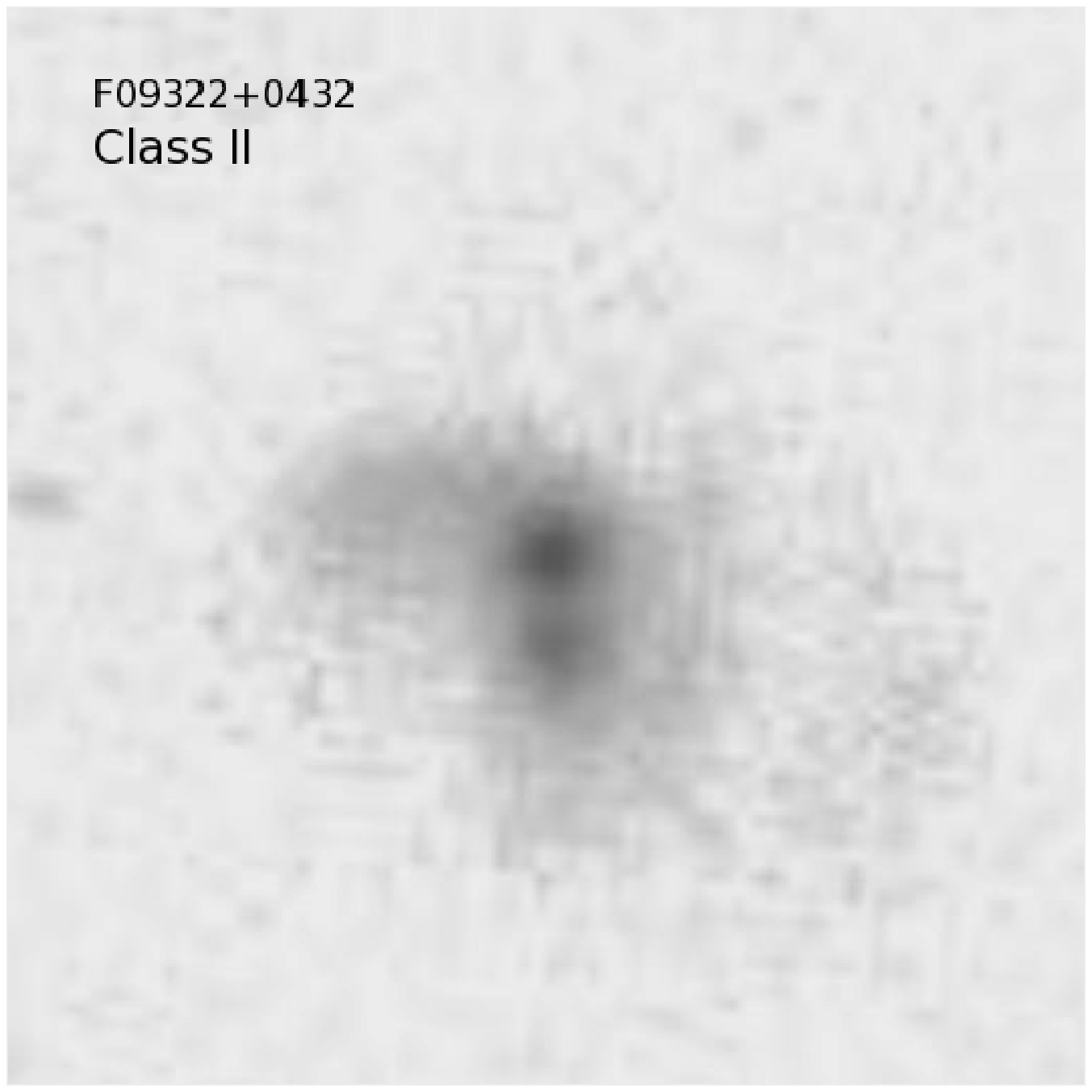}\\
\centering \includegraphics[width=40mm]{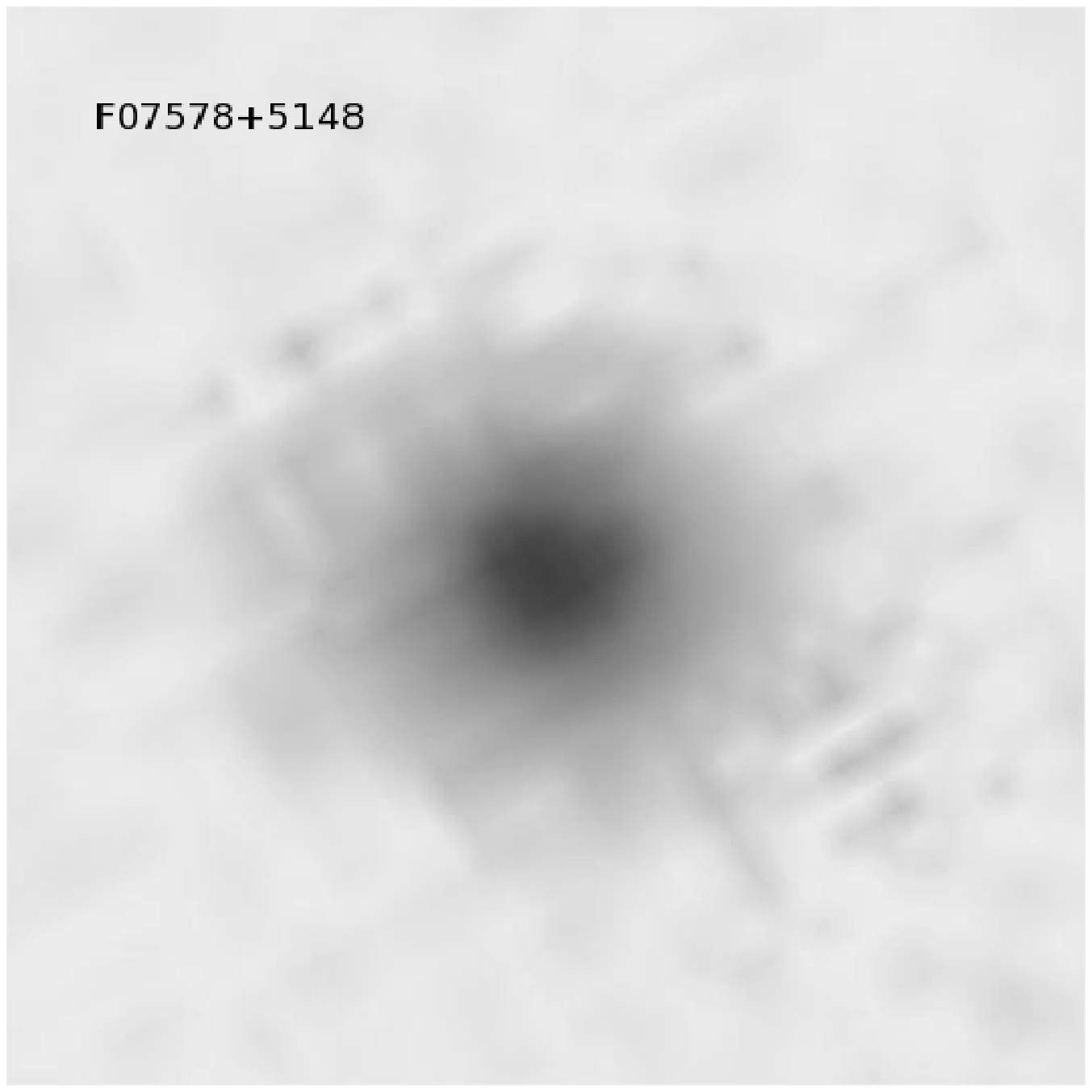}
\centering \includegraphics[width=40mm]{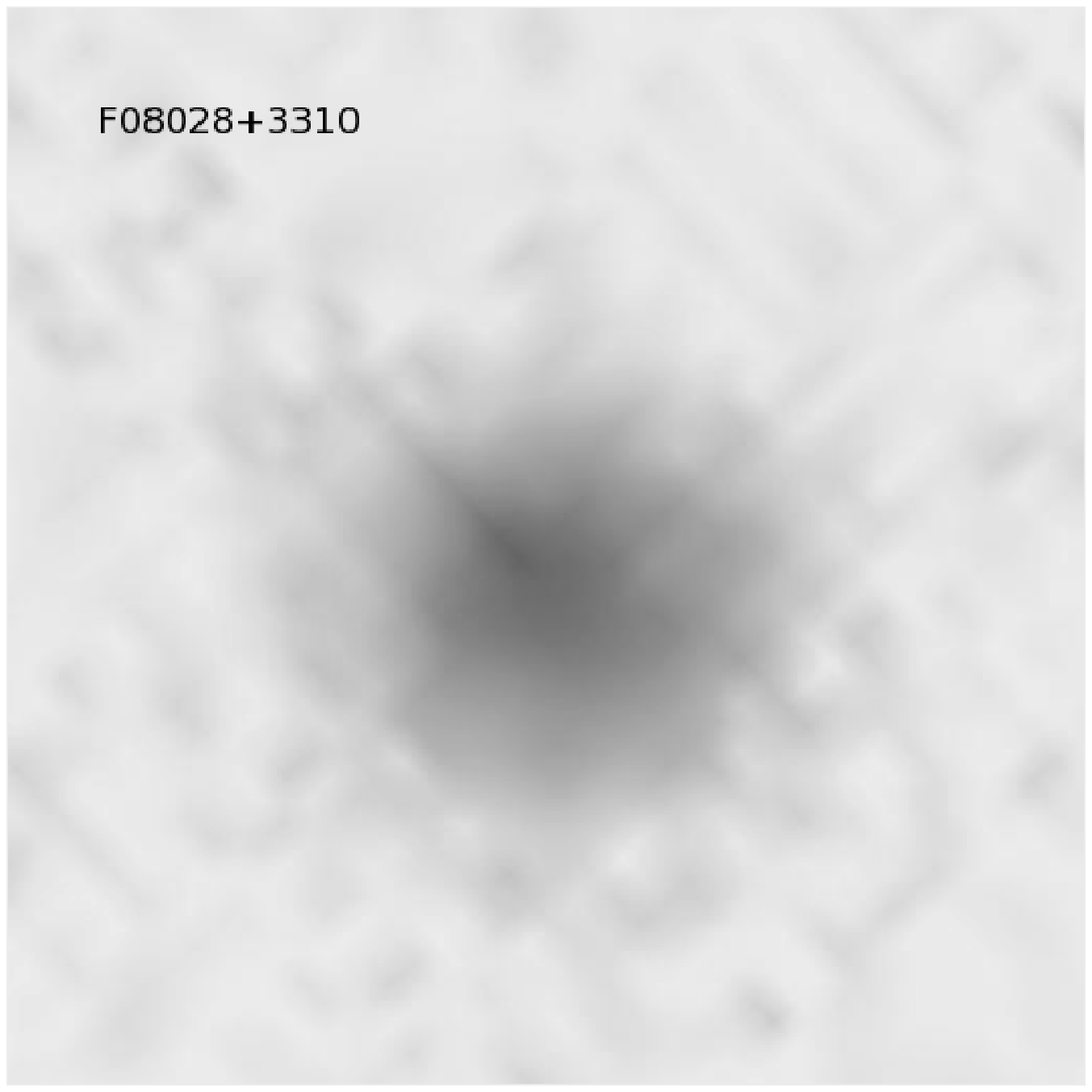}
\centering \includegraphics[width=40mm]{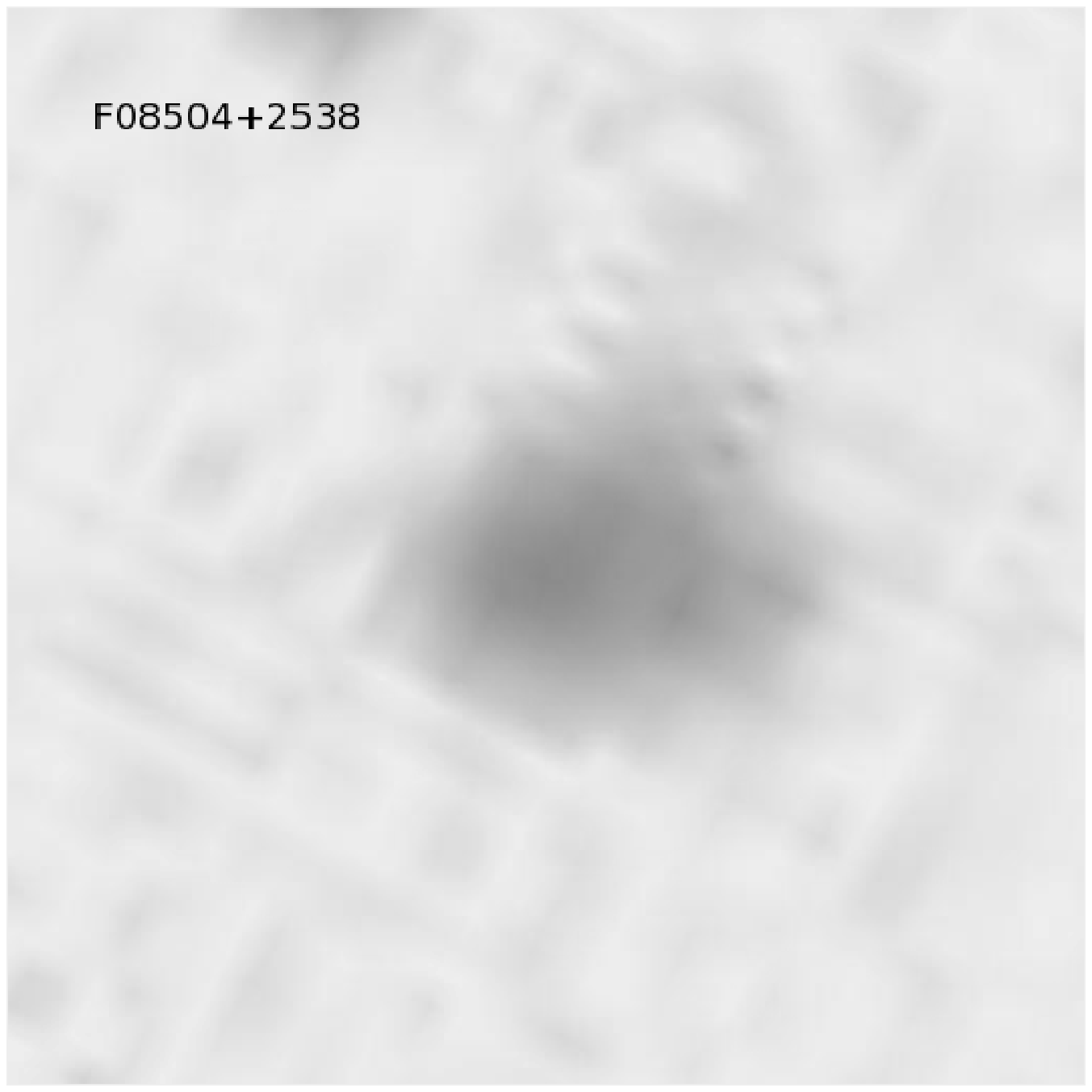}
\centering \includegraphics[width=40mm]{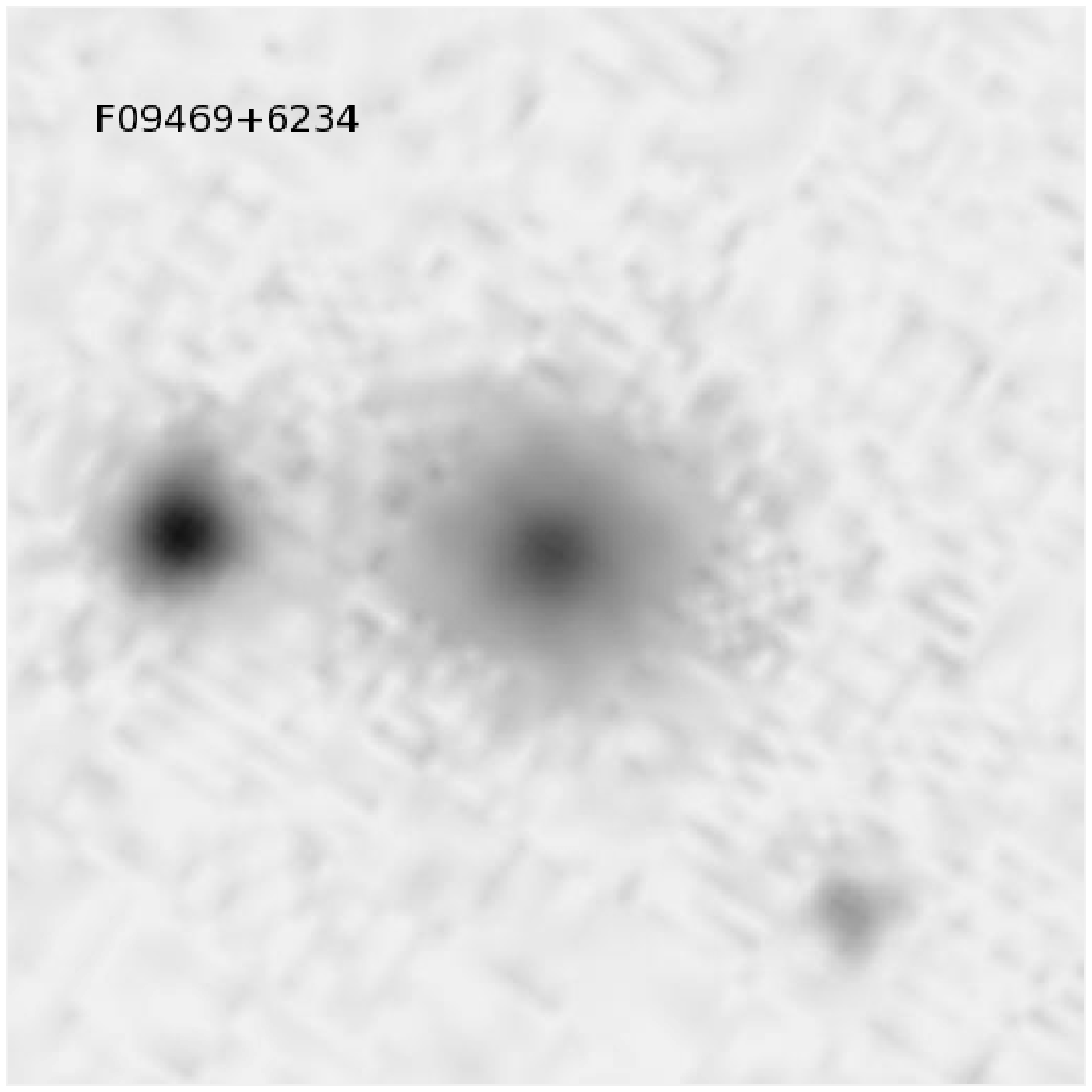}\\
\centering \includegraphics[width=40mm]{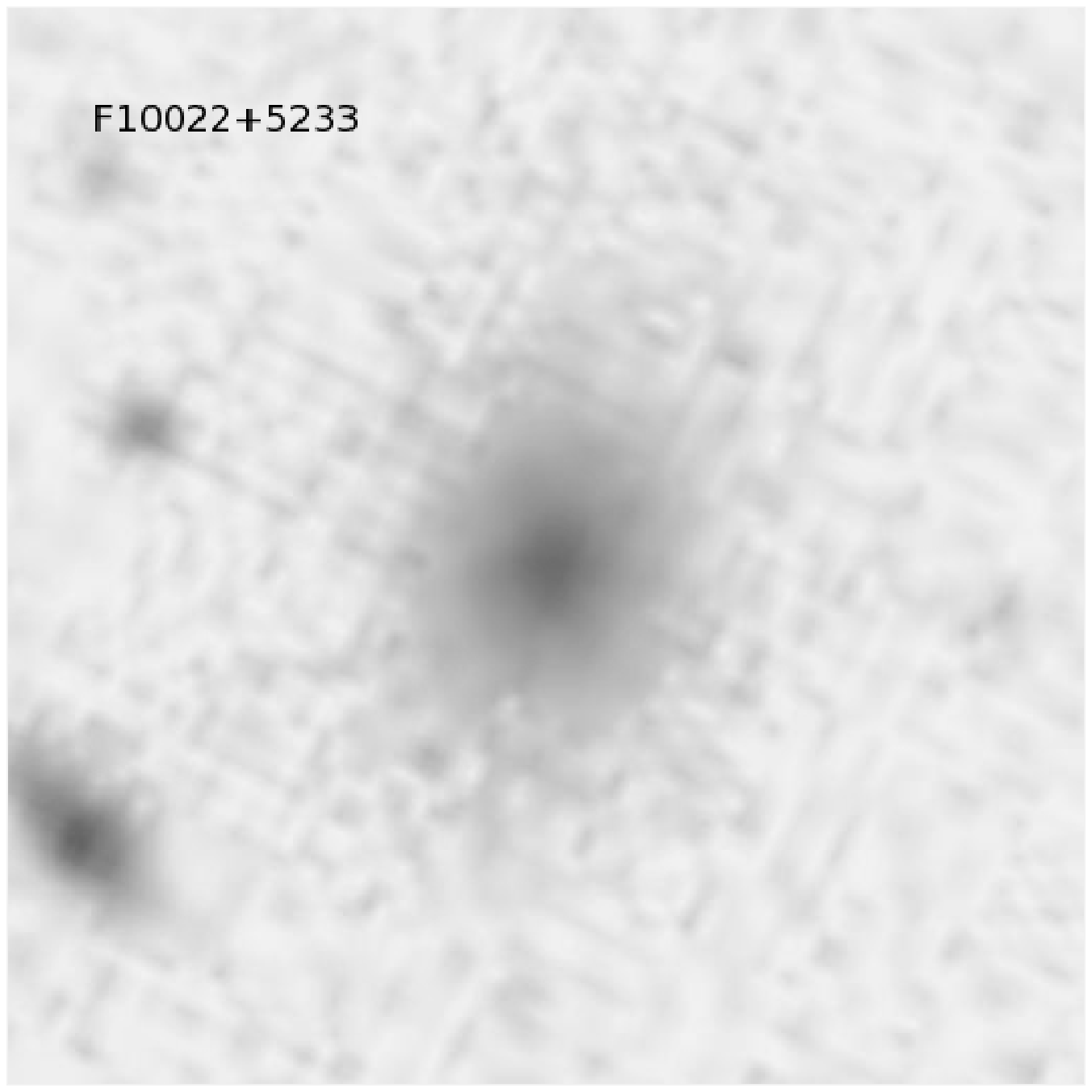}
\centering \includegraphics[width=40mm]{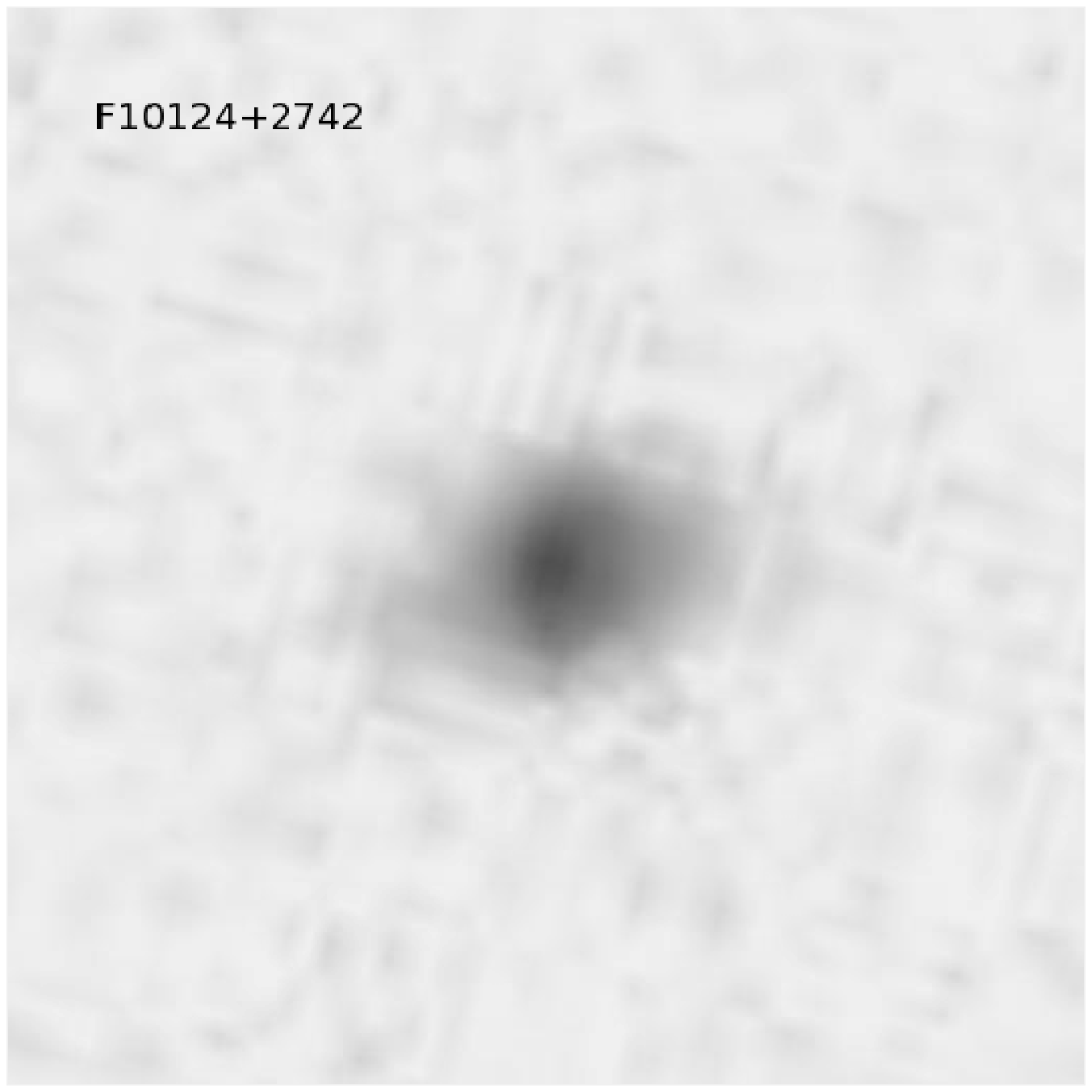}
\centering \includegraphics[width=40mm]{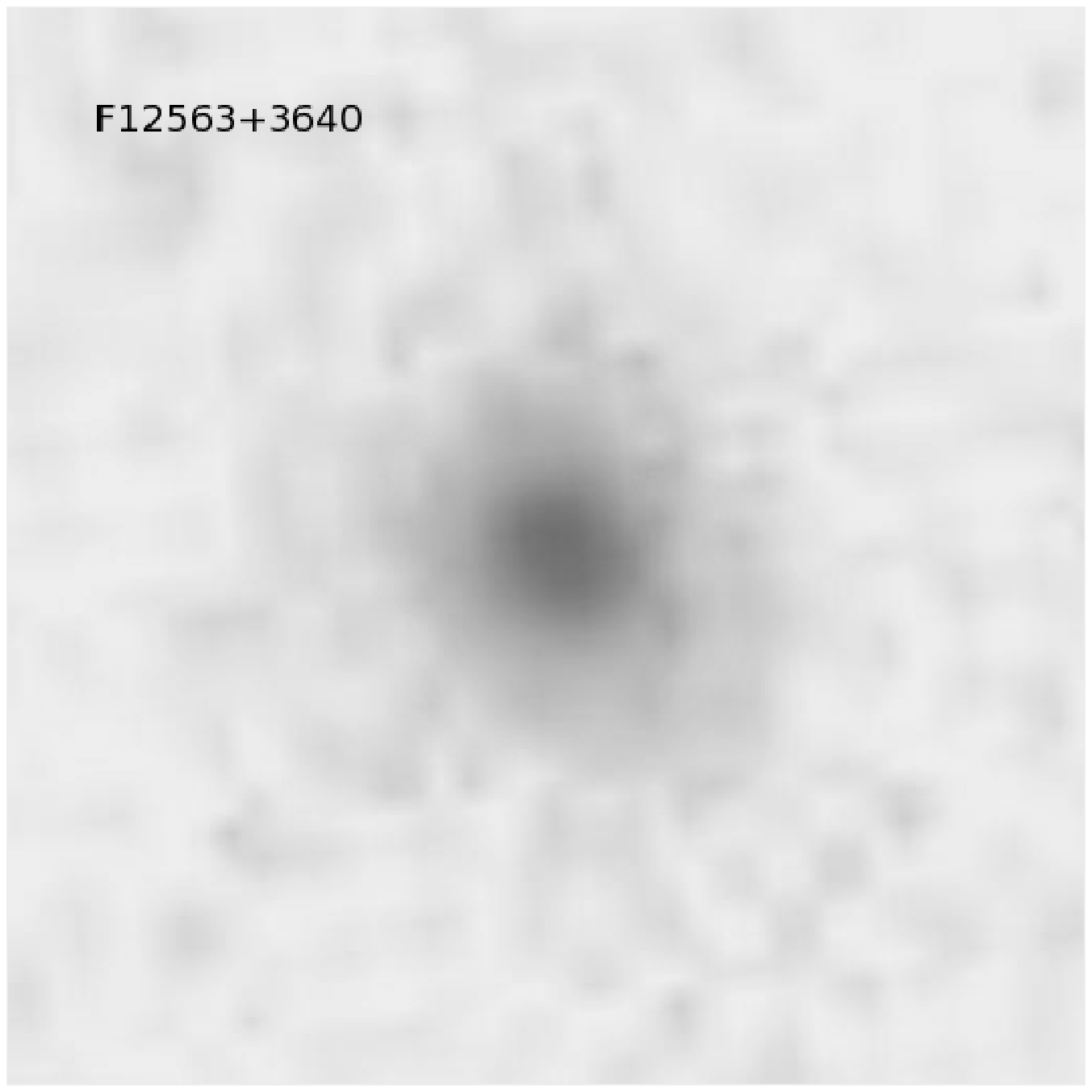}
\centering \includegraphics[width=40mm]{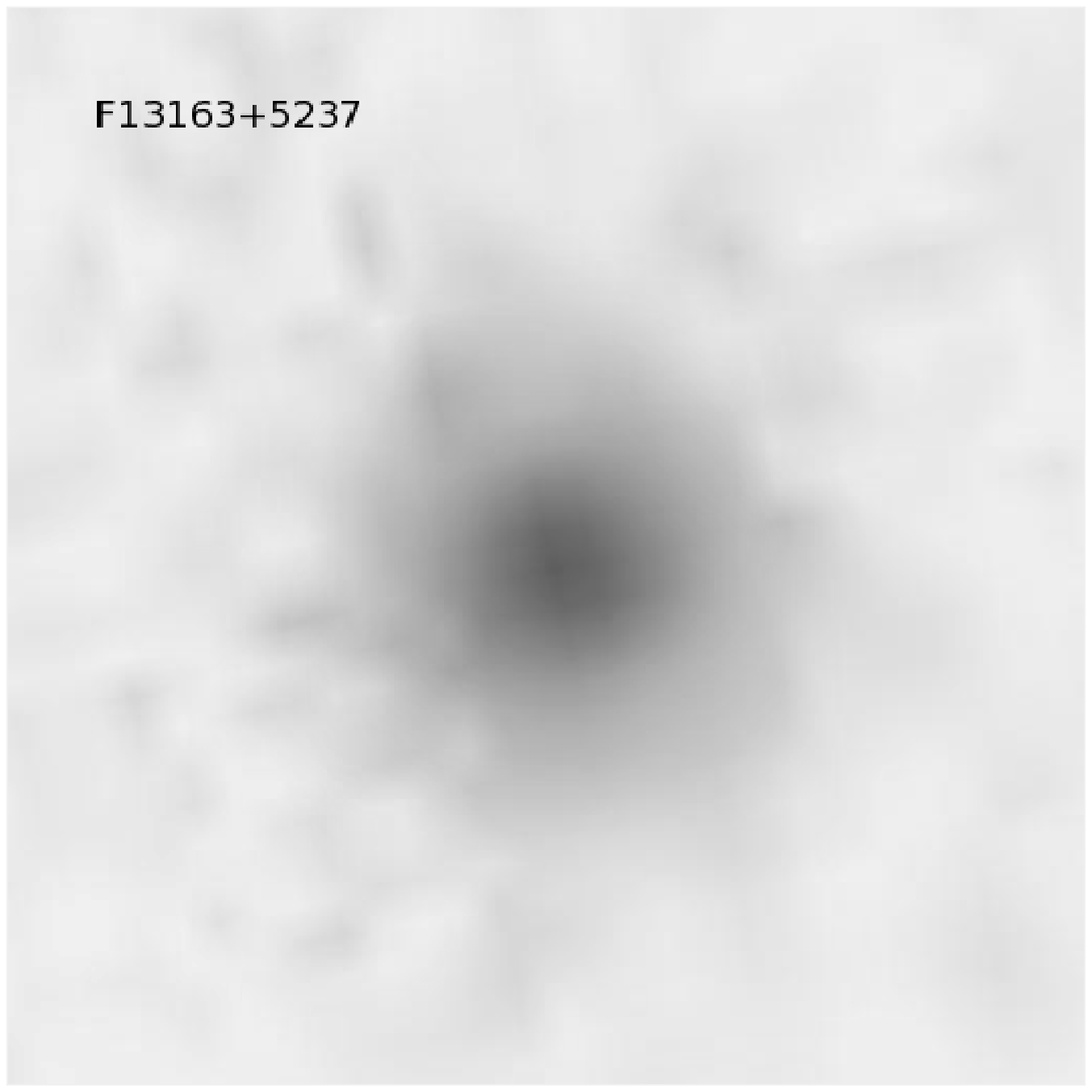}\\
\caption{Examples of SDSS images of ULIRGs with (the upper eight ) or
  without (the lower eight) obvious interacting features in our
  sample. {The interaction classifications are labeled for the
    upper eight ULIRGs.}}
\label{image}
\end{figure*}

\subsection{The optical images of ULIRGs}
We examine the images of these {308} ULIRGs by using the SDSS DR6
Image List Tool. Due to the limited resolution of the SDSS image, we
can only mark the ULIRGs with an obvious interaction feature.
{ULIRGs are classified into} \citep[see][]{vks02} {class} I with one
nucleus but with tail features; {class} II have two identified
nuclei and well-developed tidal tails or/and bridges; {class} III
have two close or even overlapped nuclei, and their redshifts
(almost in all cases, only one source in this system has spectral
redshift measurements, and another/others have only {SDSS}
photometric redshift data) are consistent. Examples are shown in
Figure~\ref{image}. About {56\%} of the ULIRGs in our total sample
show obvious interaction features. The nearby ULIRGs show more
clearly the interaction features, with the percentage of about
{92\%} for $z$$<0.1$, {84\%} for $z$$<0.15$, and {79\%} for
$z$$<0.2$. The selection effect is obvious in the classification of
interaction features of the ULIRGs, {because we can see the tails
and two interacting galaxies at
  the maximum redshift about $z$ = 0.3.} Objects at various merger
stages appear in our ULIRG sample. Some of them are still widely
separated, and some are advanced mergers. {The minimum
  separation between two nuclei in our sample is about 2$^{\prime\prime}$.}
{The imaging property is classified for each source in
  Table~1.}

Optical spectra of all {308} ULIRGs are available in the SDSS
archive. This is the largest sample of ULIRGs with optical spectra.

\subsection{Optical spectra of ULIRGs}

We want to identify Type I ULIRGs and NL ULIRGs from this sample.
{Only the spectra with signal-to-noise (S/N) $>$ 3 are considered in
this
  study.} After excluding the sources with no obvious emission lines
in their spectra, we found {209} sources with emission lines, {62}
of which have broad-line components and are Type I ULIRGs.

\begin{figure*}
\centering \includegraphics[width=75mm]{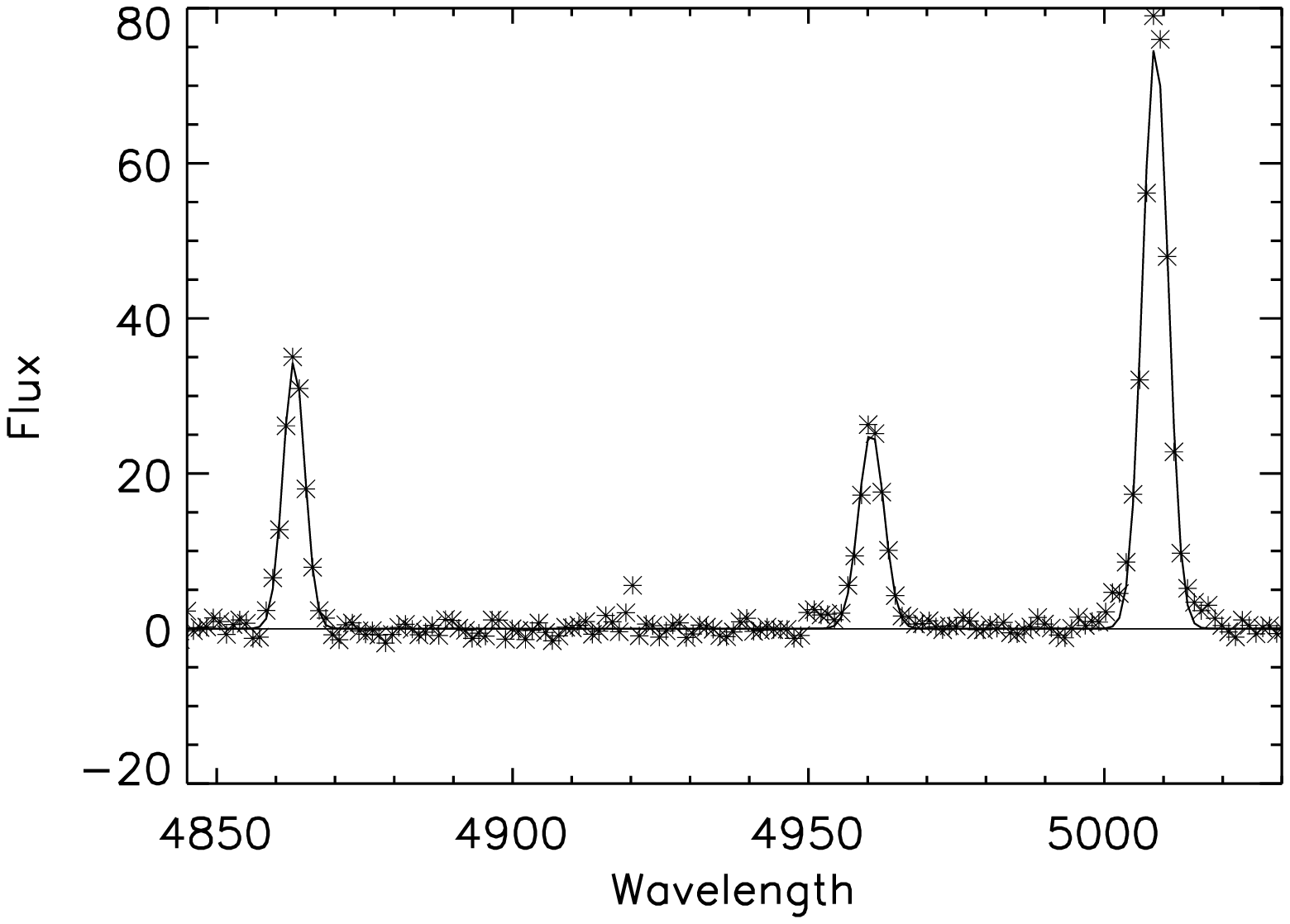}
\centering \includegraphics[width=75mm]{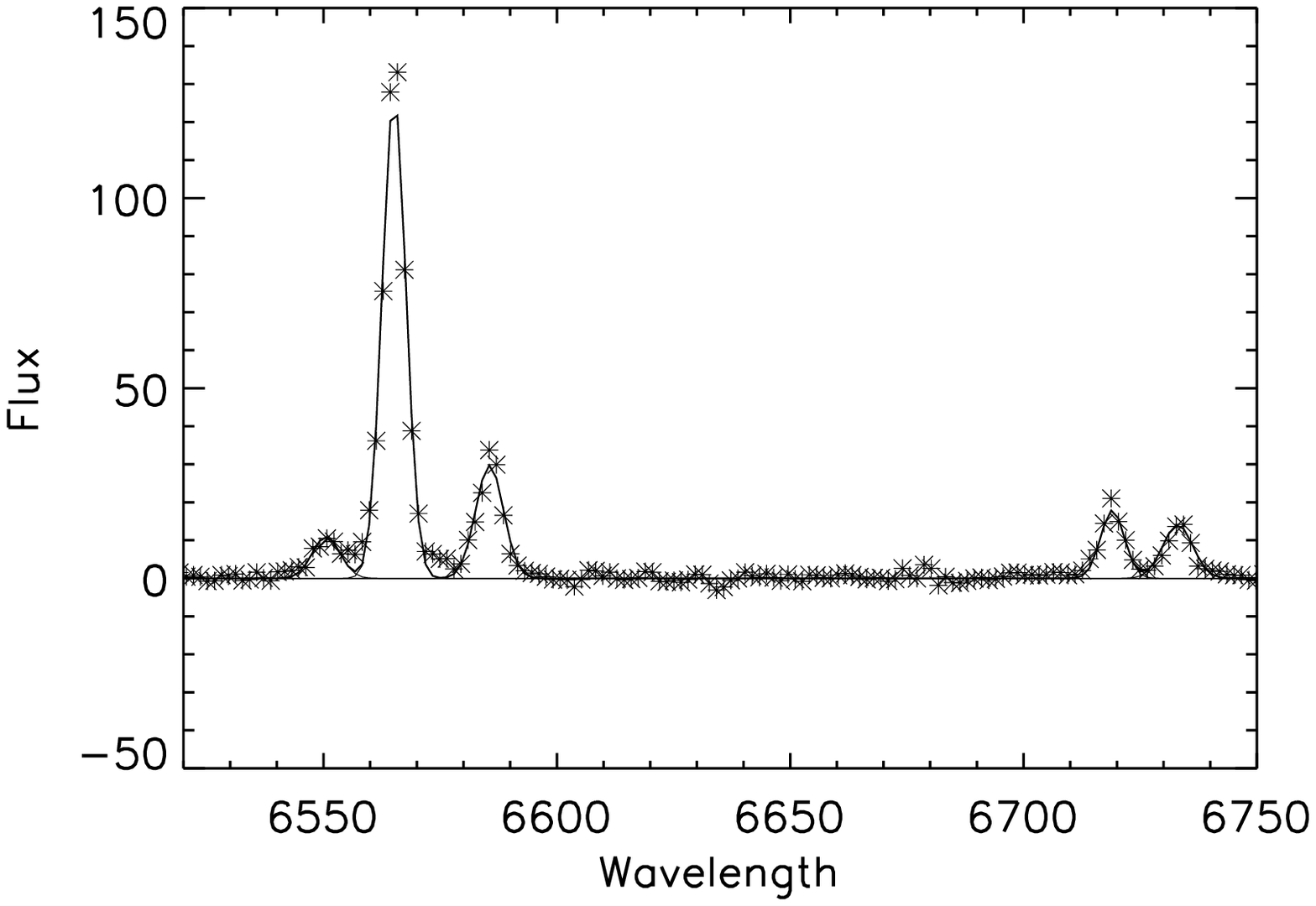}\\
\centering \includegraphics[width=75mm]{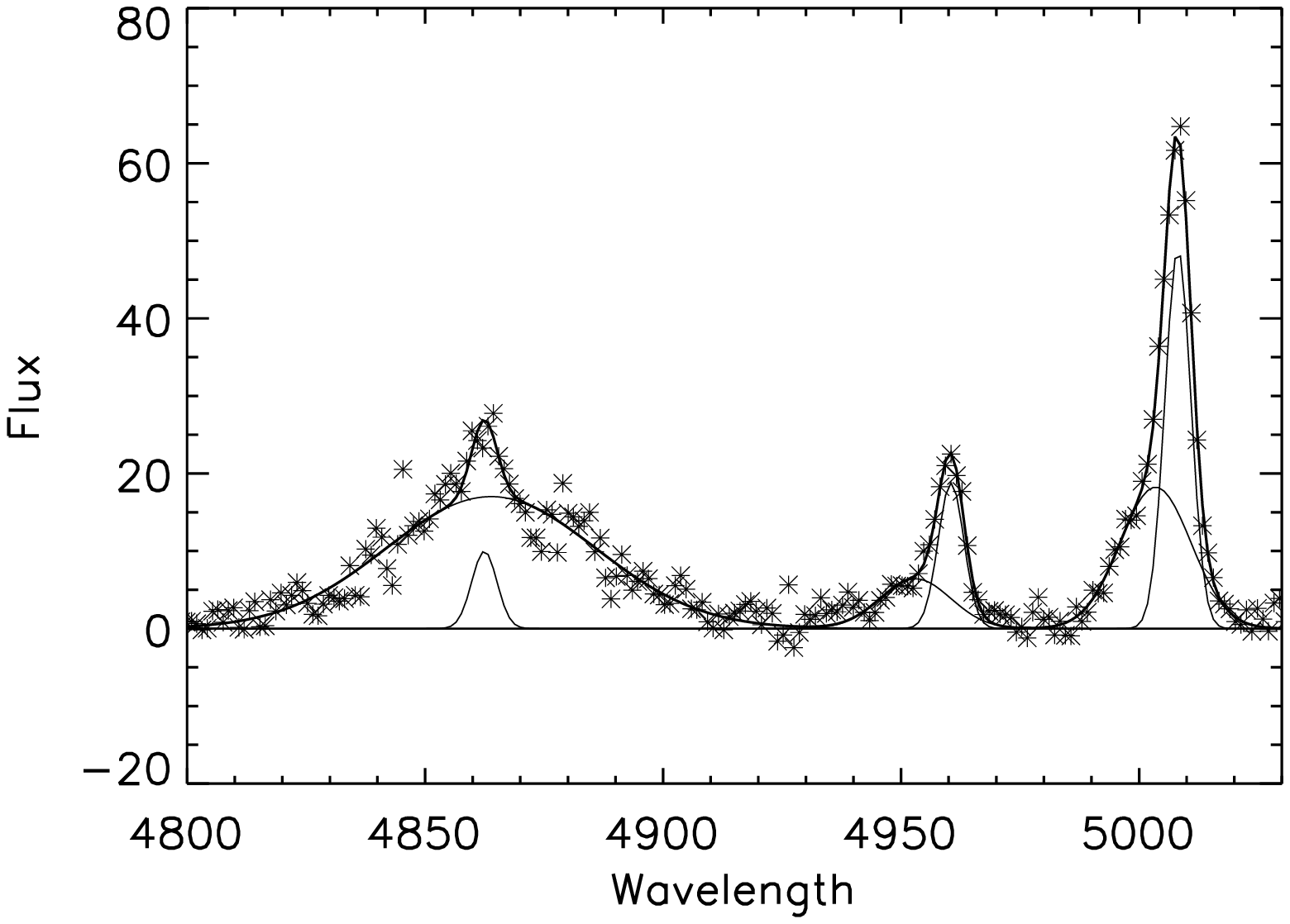}
\centering \includegraphics[width=75mm]{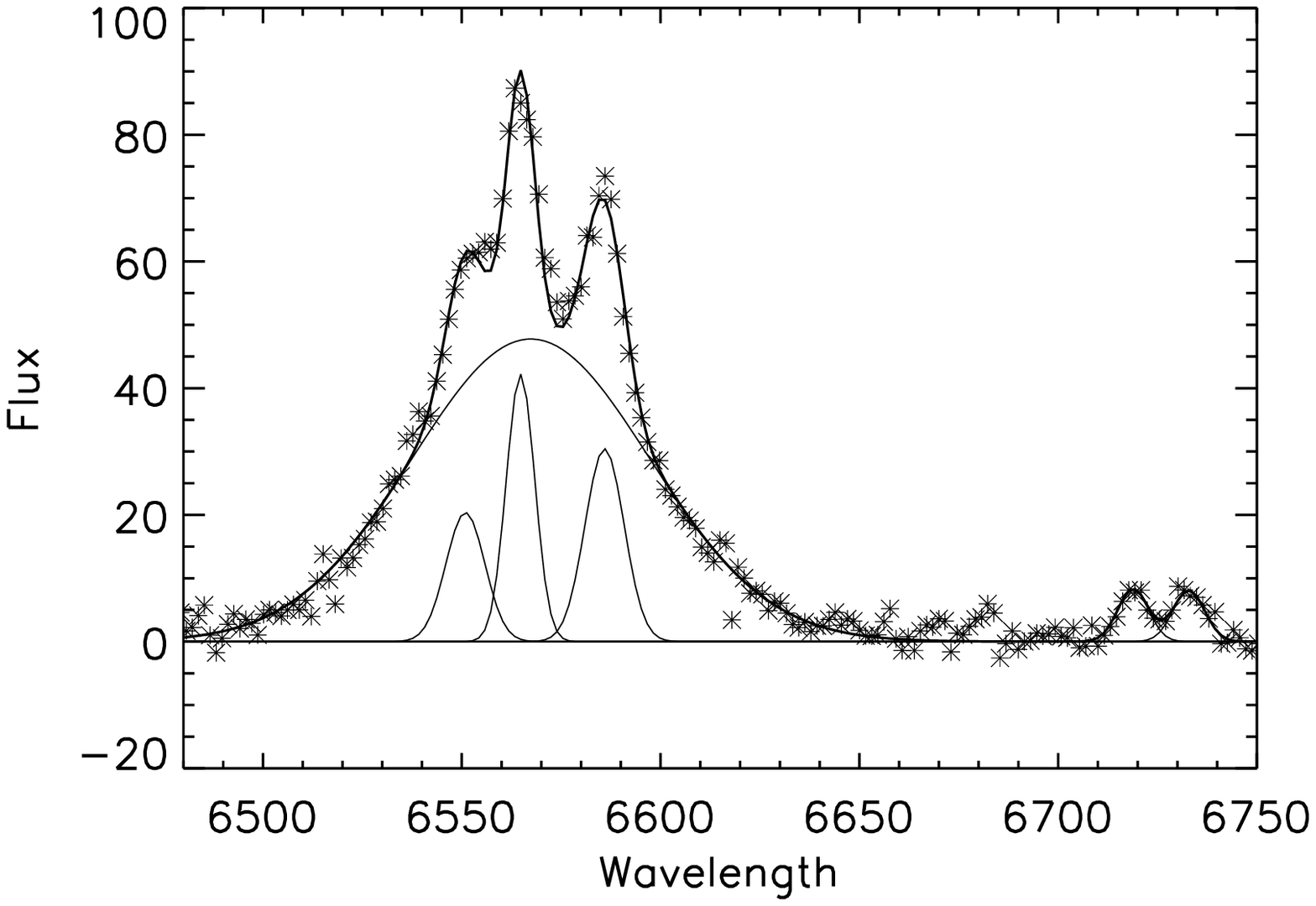}\\
\caption{Examples of our fitted SDSS spectra for the H$_{\beta}$ and
  H$_{\alpha}$ regions for a NL ULIRG (two upper panels) and a Type I
  ULIRG (two lower panels)}
\label{exam1}
\end{figure*}

SDSS spectra were processed as follows: first, we correct the
Galactic extinction and redshift effects, and then subtract the
{Fe$_{\rm II}$} emissions from the continuum by use of the optical
iron template from \citet{bg92} in the wavelength range 4250 {\AA}
$<\lambda<$ 7000 {\AA}.  Second, we fit the spectra using the Mpfit
package in IDL which is based on the Levenberg$-$Marquardt method.
The continuum emission of ULIRGs comes from central AGN and host
galaxy, and is often modified by {the intrinsic} dust extinctions.
These effects should be considered for {the determination of} the
continuum flux. A local power law is used for fitting the continuum.
After the subtraction of the fitted continuum emission from the
spectra, we fit the emission lines. For NL ULIRGs, we use a single
Gaussian profile to fit the H$_{\alpha}$, H$_{\beta}$, [O$_{\rm
III}$]4959, 5007, [N$_{\rm II}$]6548, 6583, [S$_{\rm II}$]6716, 6731
, and [O$_{\rm I}$]6300 emission lines. When a single Gaussian
cannot fit the profile of emission lines very well, double Gaussians
are {used} to obtain the line flux. For Type I UILRGs, we use two
Gaussian components to fit H$_{\alpha}$ and H$_{\beta}$. If the
[O$_{\rm III}$] emission lines can not be well fitted by a single
Gaussian, two Gaussian components are also {used}. Examples of the
fitted spectra are given in Figure~\ref{exam1}.


\section{NL ULIRGs}
The emission-line properties of 147 NL ULIRGs in our sample can be
obtained {and are} listed in Table~1. We first classify them using
the Baldwin$-$Phillips$-$Terlevich (BPT) diagram \citep[][ see
Figure~\ref{bpt}]{bpt81}.

The Balmer decrement method is often used to evaluate the {intrinsic
reddening} effect. {In previous works}
\citep[e.g.][]{kgkh06,vks99,vsk99}, H$_{\alpha}$/H$_{\beta}$ $=$
2.85 was used for galaxies whose emissions are dominated by star
formation, and H$_{\alpha}$/H$_{\beta}$ $=$ 3.1 is used for galaxies
dominated by the AGN. But before we classify the object through the
BPT diagrams, we can not assess which source {is} dominated by the
star formation or the AGN. {In this work, we adopt }the intrinsic
ratio H$_{\alpha}$/H$_{\beta}$ $=$ 3.1, and the reddening curve of
\citet[][hereafter CCM89]{ccm89}, and also assume
$R_v$~=~$A_v$/$E(B-V)$~=~3.1 to do the optical classification. In
order to examine the influence of the intrinsic ratio of
H$_{\alpha}$/H$_{\beta}$, we also used H$_{\alpha}$/H$_{\beta}$ $=$
2.85 to do the classification again, and found that the
classification of only one source (F09444+1019) is not consistent
{with the case of assuming H$_{\alpha}$/H$_{\beta}$
  $=$ 3.1}.

The intrinsic flux at wavelength $\lambda$ can be expressed as
\begin{equation}
  {I(\lambda)=F(\lambda)\times 10^{cf(\lambda)} }.
\end{equation}
The relation between the intrinsic and observed line ratio of
H$_{\alpha}$ and H$_{\beta}$ is
\begin{equation}
  {I(H_{\alpha})/I(H_{\beta})=10^{c(f(H_{\alpha})-f(H_{\beta}))}\times F(H_{\alpha})/F(H_{\beta})}
  ,
\end{equation}
where $f(\lambda)$ is related to the generalized CCM89 reddening
curve, {$I(\lambda)$ is the intrinsic flux, $F(\lambda)$ is the
  observed flux, }and $c$ is the value of the Balmer extinction. We
estimated $c$ from the observed line ratio and the reddening curve,
after the intrinsic line ratio is assumed.
\begin{figure}
\centering \includegraphics[angle=270,width=75mm]{NII_BPT.ps}\\
\centering\includegraphics[angle=270,width=75mm]{SII_BPT.ps}\\
\centering\includegraphics[angle=270,width=75mm]{OI_BPT.ps}\\
\caption{BPT diagrams for 147 NL ULIRGs. The Balmer decrement method
  has been used to derive the extinction corrected flux of each
  emission line. Upper panel: the [N$_{\rm II}$]/H$_{\alpha}$ vs.
  [O$_{\rm III}$]/H$_{\beta}$ diagnostic diagram. The solid line is from
  \citet{kgkh06}, the dashed line is from \citet{kht+03}. Middle panel:
  the [S$_{\rm II}$]/H$_{\alpha}$ vs. [O$_{\rm III}$]/H$_{\beta}$ diagnostic
  diagram. The two lines are from \citet{kgkh06}. Lower panel: The
  [O$_{\rm I}$]/H$_{\alpha}$ vs. [O$_{\rm III}$]/H$_{\beta}$ diagnostic diagram. The
  two lines are from \citet{kgkh06}.}
 \label{bpt}
\end{figure}
We use the new classification scheme made by \citet{kgkh06} for the
classification.

The 147 NL ULIRGs are classified as five types: 29 star-forming
galaxies, 62 composite galaxies \citep[which are likely to contain
metal-rich stellar populations, plus AGN, see][]{kgkh06}, 34 Seyfert
galaxies, 6 LINERs, and 16 ambiguous galaxies, which are classified
as one type in one or two diagrams but another type in the other
diagram(s). Together with the {62} Type I ULIRGs, the percentage for
AGNs is about {$78\%$} if we regard that the composite galaxies also
contain AGNs. The AGN percentage {becomes} {49\%} if we only
consider the Seyfert galaxies, LINERs, and Type I ULIRGs. The AGN
percentage of ULIRGs increases with the infrared luminosity,
consistent with {the
  results in} previous works
\citep[e.g.][]{vks+95,vsk97,wzxd98,cww+06}. The percentage ({only
consider the Seyfert galaxies, LINERs, and Type I ULIRGs
  as AGNs}) is {71\%} for ULIRGs with $L_{\rm IR}$ $>$ 12.3
$L_{\odot}$, and {89\%} for $L_{\rm IR}$ $>$ 12.4 $L_{\odot}$.

A tight correlation between far-infrared and radio luminosities,
covering about four orders of magnitude in the $L_{\rm IR}$, has
been found for infrared-selected galaxies
\citep[e.g.][]{hsr85,yrc01}. The non-thermal radiation from the
relativistic electrons related to the supernova remnants is believed
to be the main reason of this relation. Here, we verify this
relation {using our ULIRG
  sample. The FIRST flux density was used to calculate the radio
  luminosity. In our sample, 214 ULIRGs have FIRST counterparts,
  including 118 NL ULIRGs and 45 Type I ULIRGs.}

\begin{figure}
\centering \includegraphics[angle=270,width=75mm]{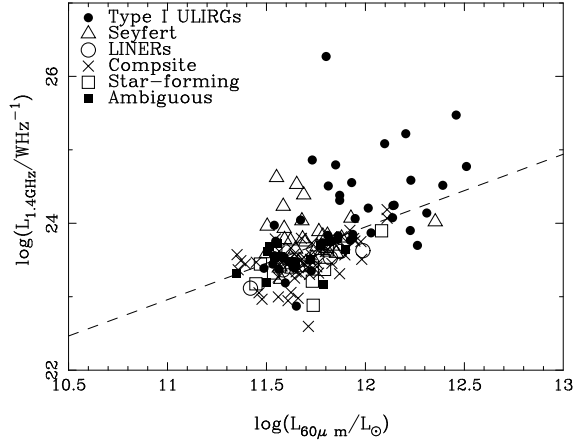}
\caption{1.4 GHz radio luminosity vs. \emph{IRAS} 60 $\mu$m
luminosity
  for our NL ULIRGs and Type I ULIRGs. The dashed line is the best fit
  given by \citet{yrc01} from 1809 infrared galaxies.}
\label{lirradio}
\end{figure}

\begin{figure}
\centering \includegraphics[angle=270,width=75mm]{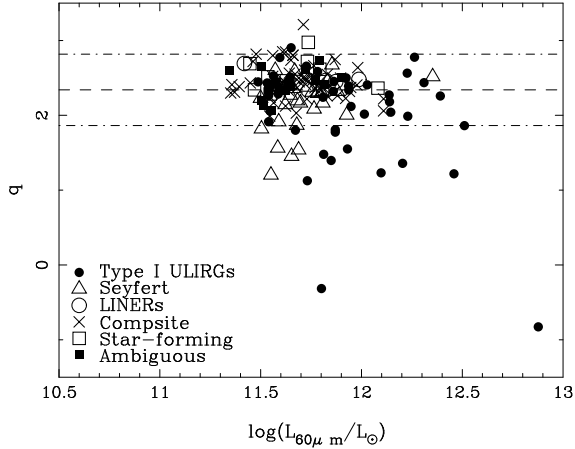}
\caption{The $q$-values plotted against the \emph{IRAS} 60 $\mu$m
  luminosities for the same sample used in Figure~\ref{lirradio}.
  {The dashed line marks the average value of $q$ = 2.34 obtained
    by \citet{yrc01} from a sample of 1809 infrared galaxies. The
    two dot-dashed lines represent a three times larger radio (lower line) or IR
    (upper line) flux density than the expected values from the linear
    radio$-$FIR relation of \citet{yrc01}. }}
\label{qqlir}
\end{figure}

The luminosities at 60 $\mu$m and at 1.4 GHz radio band are obtained
using the following formulae \citep[][]{yrc01}:
\begin{equation}
  {{\rm log}L_{\rm 1.4GHz}(\rm  W~Hz^{-1})=20.08+2{\rm log}D+{\rm log}S_{1.4 \rm GHz}} ,
\end{equation}
\begin{equation}
  {{\rm log}L_{\rm 60\mu m}(L_{\odot})=6.014+2{\rm log}D+{\rm log}S_{60\mu m}} ,
\end{equation}
where $D$ is the luminosity distance in Mpc, $S_{1.4{\rm GHz}}$ and
$S_{60{\rm \mu m}}$ are flux densities in units of Jy. The
radio$-$FIR relation of our ULIRGs is shown in
Figure~\ref{lirradio}. To examine the deviation of these ULIRGs from
the linear relation, we also calculated the $q$ parameter
\citep[see][]{cah91,yrc01},
\begin{equation}
  q={{\rm log}(\frac{2.58S_{\rm 60\mu m}+S_{100\mu m}}{\rm 2.98Jy})-{\rm log}(\frac{S_{\rm 1.4GHz}}{\rm Jy})} .
\end{equation}
A plot of $q$ versus $L_{60 {\rm \mu m}}$ is shown in
Figure~\ref{qqlir}. The radio excess objects (i.e., {the objects
  having 3 times larger radio flux density than the expected values
  from the linear radio$-$FIR relation of \citet{yrc01}}) are
{AGNs}, either Type I ULIRGs, or Seyfert galaxies. {Therefore}, the
radio$-$FIR relation originates from the starburst-related
non-thermal radiation, and the radio excess objects are due to the
AGN-related radio emission \citep{rn97}.

\section{Type I ULIRGs}

\begin{figure*}
\centering \includegraphics[width=75mm]{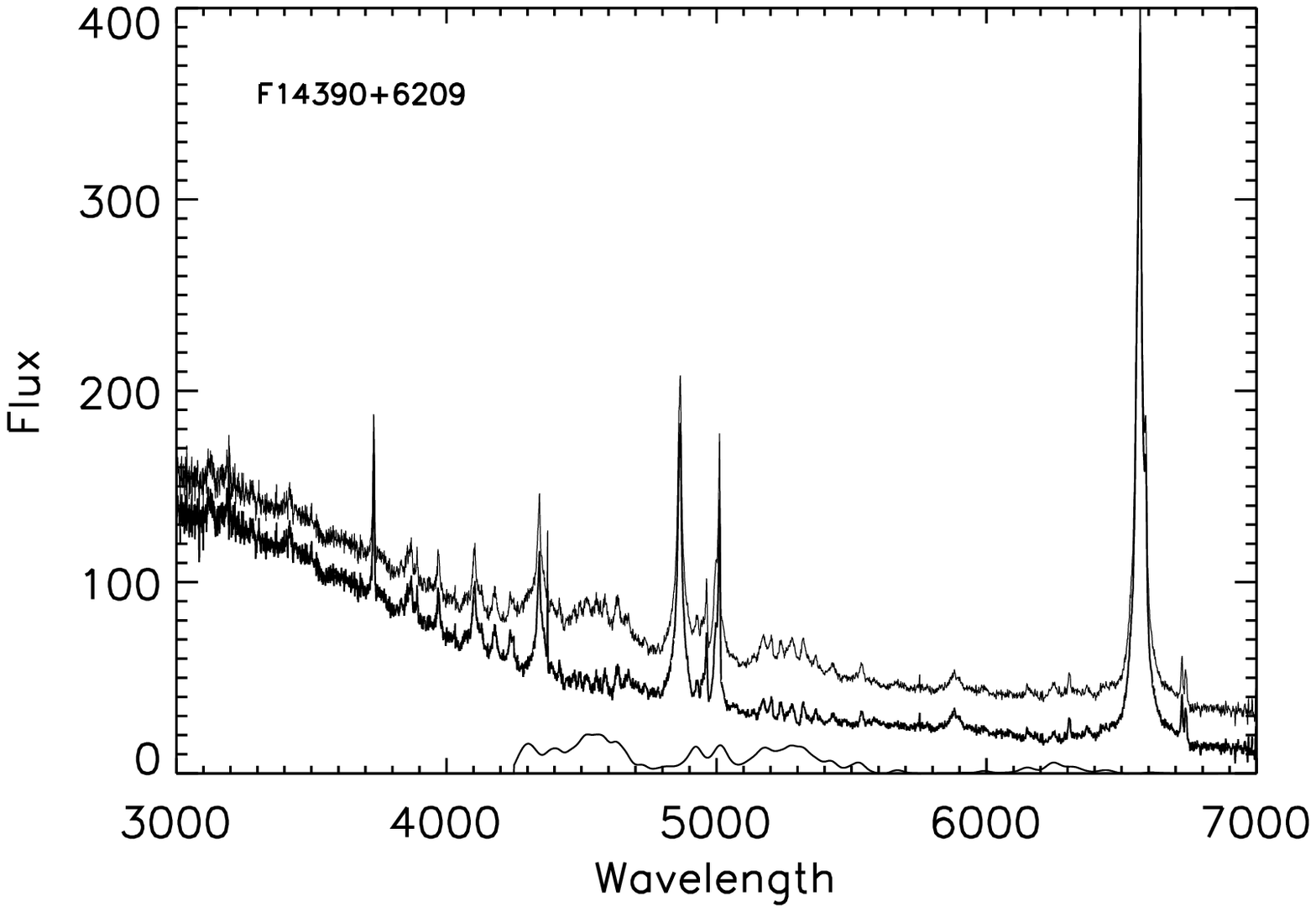}
\centering \includegraphics[width=75mm]{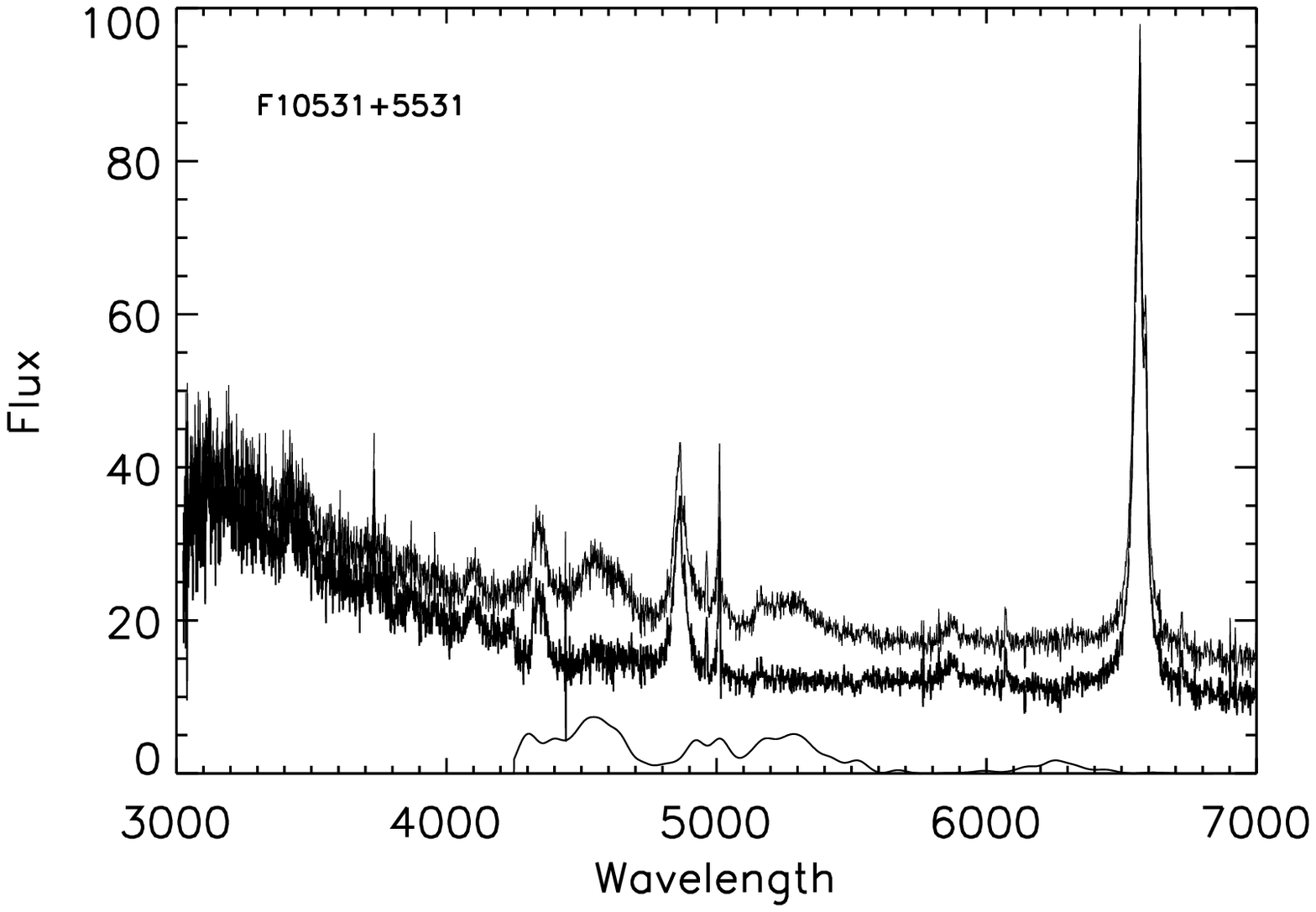}\\
\centering \includegraphics[width=75mm]{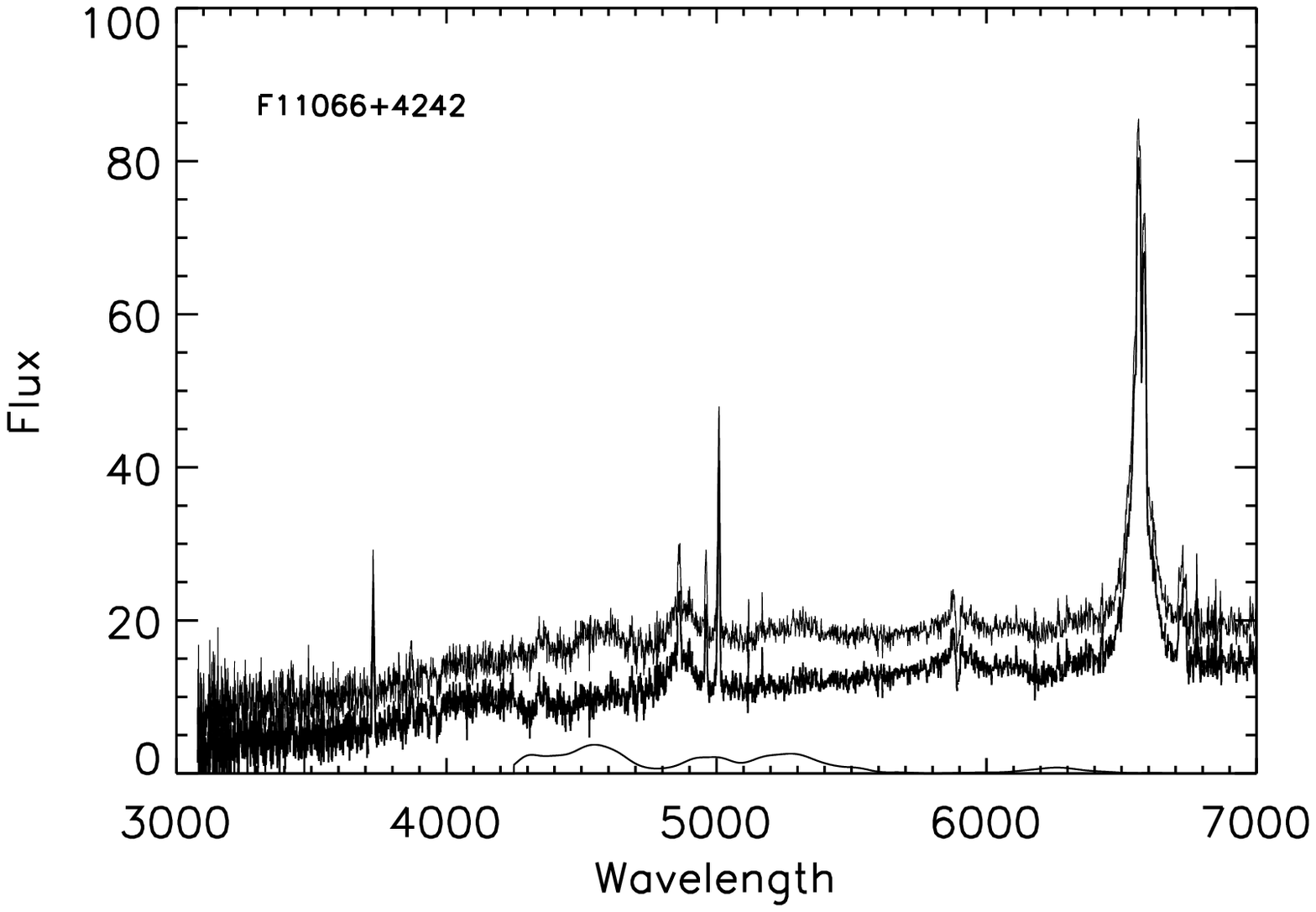}
\centering \includegraphics[width=75mm]{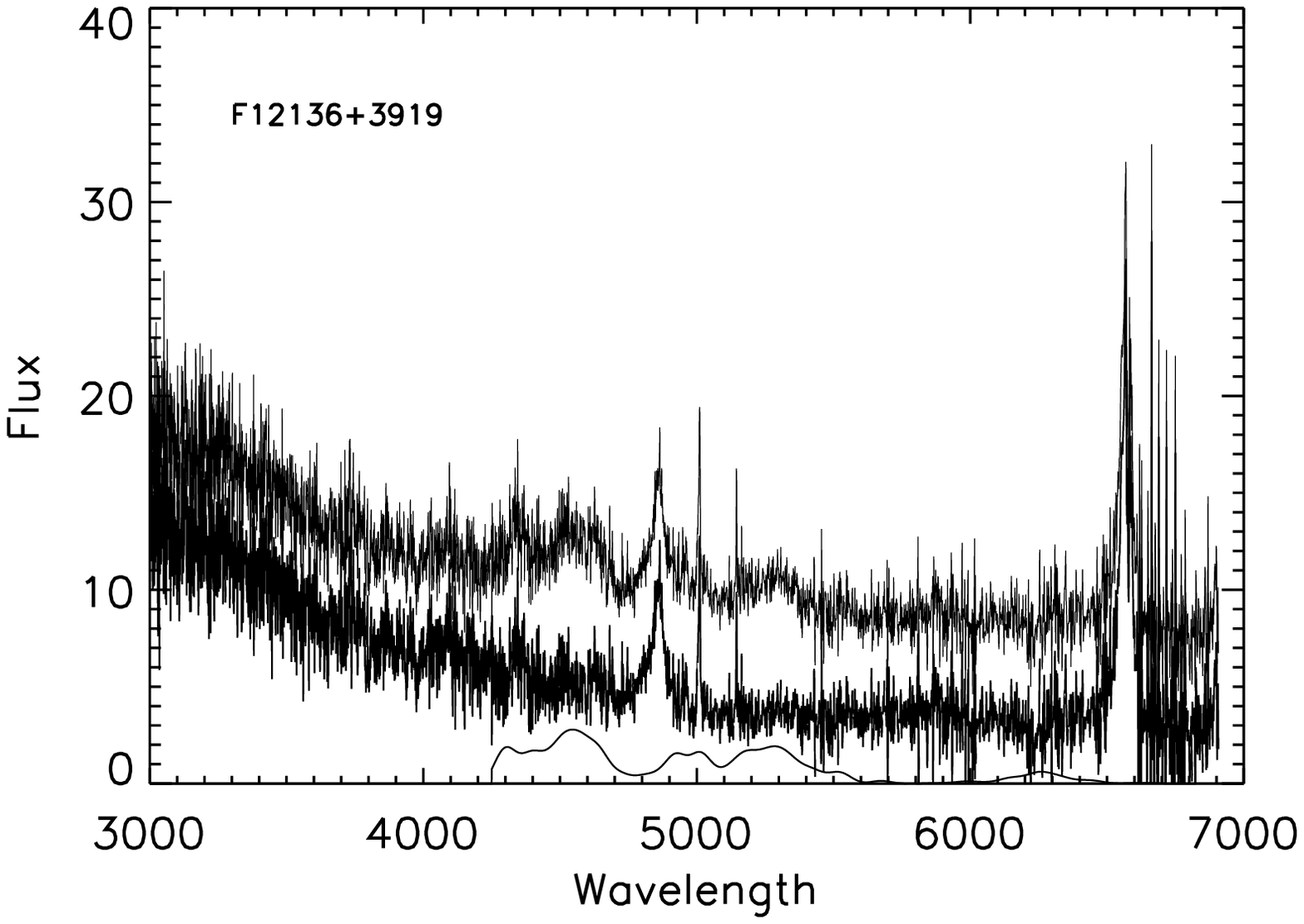}\\
\caption{Examples of SDSS spectra of Type I ULIRGs in our sample. In
  each panel, the top curve is the extinction and redshift-corrected
  spectra.  The middle curve is the spectra after the subtraction of the
Fe$_{\rm II}$ emissions, which was shifted downward. The bottom
curve is the model
  for Fe$_{\rm II}$ emissions. }
\label{spec}
\end{figure*}

In our ULIRG sample, there are {62} Type I ULIRGs. We can explore
their properties of central BHs. Examples of their SDSS spectra are
shown in Figure~\ref{spec}. The parameters of these Type I ULIRGs
are listed in Table 2.

\subsection{Black hole masses of Type I ULIRGs}
We assume that the motion of the gas moving around the central BH is
dominated by the gravitation force and that the gas of broad
emission line region (hereafter BLR) is virialized
\citep[e.g.][]{pw99,pw00}. The BH mass can then be expressed as
$M{_{\rm BH}=\frac{R_{BLR}V^2}{G}}$, and the realistic formula given
by \citet{ksn+00} is:
\begin{equation}
 {\frac{M_{\rm BH}}{M_{\odot}}=1.464\times10^5(\frac{R_{\rm BLR}}{{\rm lt-days}})}(\frac{v_{\rm FWHM}}{10^3{ \rm km~s}^{-1}})^2 ,
\end{equation}
where $v_{\rm FWHM}$ is the FWHM of the broad emission line, and
$R_{\rm
  BLR}$ is the radius of the BLR. For our objects, the $v_{\rm FWHM}$ is
taken as the FWHM of the broad component of the H$_{\beta}$ emission
line, and the $R_{\rm BLR}$ can be estimated from the Fe$_{\rm II}$
and the Galactic-extinction-corrected continuum luminosity at 5100
${\rm
  \AA}$ , using:
\begin{equation}
  {\frac{R_{\rm BLR}}{{\rm lt-days}}=(26.4\pm4.4)[\frac{\lambda L_{\lambda}({5100{\AA}})}{10^{44}\rm erg~s^{-1}}]^{(0.61\pm0.10)}}  .
\end{equation}
The relation between $R_{\rm BLR}$ and $L_{\lambda}({\rm
5100{\AA}})$ was first found by \citet{ksn+00}, and their data were
refitted by \citet{mj02} in the same cosmology as we adopted.
\begin{figure}
\centering
\includegraphics[angle=270,width=75mm]{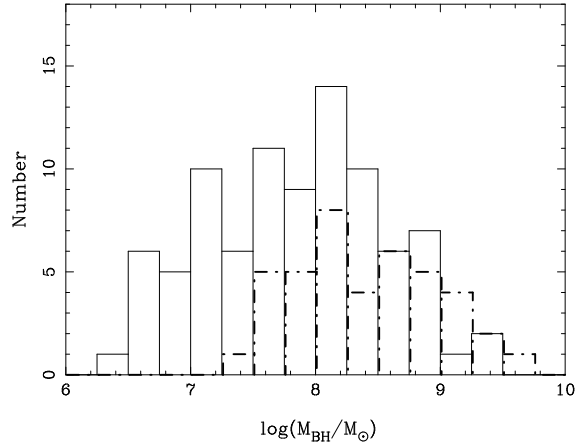}
\caption{BH mass distribution. The solid line represents the BH mass
  distribution of our sample and the Type I ULIRGs from Hao et
  al.(2005), while the dot-dashed line is for that of the PG QSOs obtained
  from Hao et al.(2005).} \label{bhmass}
\end{figure}
Here {we assume that the} central AGN dominates the continuum
emission at 5100 ${\rm \AA}$, and the contribution from stellar emission and
the {intrinsic} reddening effect can be neglected. \citet{kan+06} argued that, for
Type I ULIRGs, the central AGN dominance of the 5100 ${\rm \AA}$ emission
is reasonable.
\begin{figure}
\centering \includegraphics[angle=270,width=75mm]{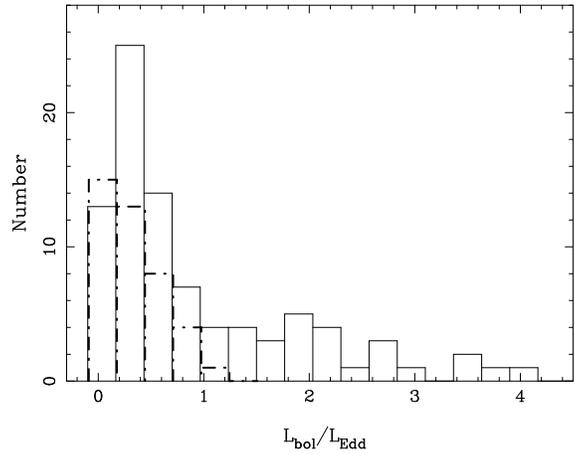}
\caption{Eddington ratio distribution of 90 Type I ULIRGs from our
  sample and \citet{hxm+05}, {while the dot-dashed line is for that of the
  PG QSOs obtained from Hao et al.(2005).} }
\label{ratio}
\end{figure}

To investigate the BH masses of Type I ULIRGs systematically, a
large sample is absolutely needed. \citet{kan+06} used eight ULIRGs
from the Type I ULIRG sample of \citet{zxm+02} to investigate this
problem, and concluded that the BH masses of Type I ULIRGs are
typically smaller than those of PG QSOs. The FWHM$_{\rm H_{\beta}}$
they used is not the broad component of H$_{\beta}$, which leads to
an underestimation of the BH mass. \citet{hxm+05} carried out a
study on Type I ULIRGs, {with a sample} mainly from \citet{zxm+02},
but added several PG QSOs. The FWHM of the H$_{\beta}$ broad
component they used was estimated in the same way as that in
{\citet{bg92}, who used
  double components to fit the H$_{\beta}$ line}. \citet{hxm+05} used
the cosmology of $H_0$ = 70 km~s$^{-1}$~Mpc$^{-1}$, $\Omega_{\rm m}$
= 0.3, and $\Omega_{\Lambda}$ = 0.7, the same as ours. We noted that
{three} of our {62} Type I ULIRGs were listed in their sample: for
F01572+0009, the BH mass in our work (log(${
  M_{\rm BH}/M_{\odot})\sim8.0}$) is consistent with their result
(log(${ M_{\rm BH}/M_{\odot})\sim7.7}$). {For F10026+4347, the
  same BH mass (log(${ M_{\rm BH}/M_{\odot})\sim7.8}$) is given in
  \citet{hxm+05} and our work.} But {for} the other source F13342+3932, the
BH mass we obtained (log(${ M_{\rm BH}/M_{\odot})\sim8.4}$) is
larger than theirs (log(${ M_{\rm BH}/M_{\odot})\sim7.4}$), mainly
because we use the much more broad component of H$_{\beta}$ from
SDSS spectra.

We combine the sample {of \citet{hxm+05} }with ours, and obtain a
large sample of {90} Type I ULIRGs. The distribution of the BH mass
of this sample is shown in Figure~\ref{bhmass}, together with that
of PG QSOs. Obviously, the BH masses of Type I ULIRGs (with a mean
value of {6.7$\times 10^7
  {M_\odot}$ for these 90 Type I ULIRGs and about 7.6$\times 10^7 {
    M_\odot}$ for our 62 Type I ULIRGs}) are systematically smaller
than those of PG QSOs (with a mean value of 2$\times 10^8 {
  M_\odot}$), consistent with the popular evolutionary scheme that
ULIRGs are in a pre-QSO phase and their central BHs are still under
growing.

{The bolometric luminosity} $L_{\rm bol}$ measures the total
luminosity associated with the AGN. We estimate the bolometric
luminosity using the formula in \citet{ksn+00}: $L{_{\rm
bol}\approx9\lambda L_{\lambda}({\rm 5100{\AA}})}$. For those ULIRGs
from \citet{hxm+05}, we directly use their data because they used
the same method to calculate the BH mass and $L_{\rm bol}$. The
distribution of the Eddington ratio of this sample is shown in
Figure~\ref{ratio}, and the mean value is about {0.92 for
  these 90 Type I ULIRGs, and 0.55 for our 62 Type I ULIRGs} , larger
than that of PG QSOs (the mean value of the Eddington ratio for them
is about 0.2).

{The intrinsic reddening effect in Type I ULIRGs is still poorly
  known, and it probably affects the estimation of the optical
  luminosity. The Balmer decrement method is often used to
  estimate the reddening effect for narrow emission galaxies, but is
  seldom used for broad emission-line galaxies. In principle, the
  observations in the X-ray band can be used to estimate the absorption
  column density and then the absorption. But for our {62} Type
  I ULIRGs, only several have been observed in the X-ray band. More work
  is still needed to understand the intrinsic reddening effect in Type
  I ULIRGs.}

\begin{figure}
\centering \includegraphics[angle=270,width=75mm]{Mbh_oiii_NLcore.ps}\\
\centering \includegraphics[angle=270,width=75mm]{Mbh_sii.ps}\\
\centering \includegraphics[angle=270,width=75mm]{Mbh_sii2.ps}

\caption{$M_{\rm BH}$$-$$\sigma$ diagrams for Type I ULIRGs.{
    The filled triangles represent the six sources with measured
    $\sigma$ in \citet{dtd06,dtd+06,dtd+07}, open triangles
    represent the seven QSOs with measured $\sigma$ in \citet{dtd+07},
    filled circles indicate the ULIRGs whose parameters are estimated in
    this work. If $\sigma$ is unavailable in reference, it was
    estimated by several methods: in the upper panel, the $\sigma$ is
    estimated from the FWHM of the [O$_{\rm III}$] NL core; in the middle panel, the
    $\sigma$ is estimated from the FWHM of the [S$_{\rm II}$] profile; in the lower panel, the
    same as the middle panel, except that for some sources,
    i.e. F11394+0108 and F17234+6228, their velocity dispersions are
    estimated from the FWHM of the [O$_{\rm III}$] NL core because the
    FWHM of the [O$_{\rm III}$] NL core is smaller than that of the [S$_{\rm II}$] profile.  The
    dashed line is the $M_{\rm BH}$$-$$\sigma$ relation from
    \citet{tgb+02}.}}
\label{mbhoiii}
\end{figure}


\begin{figure}
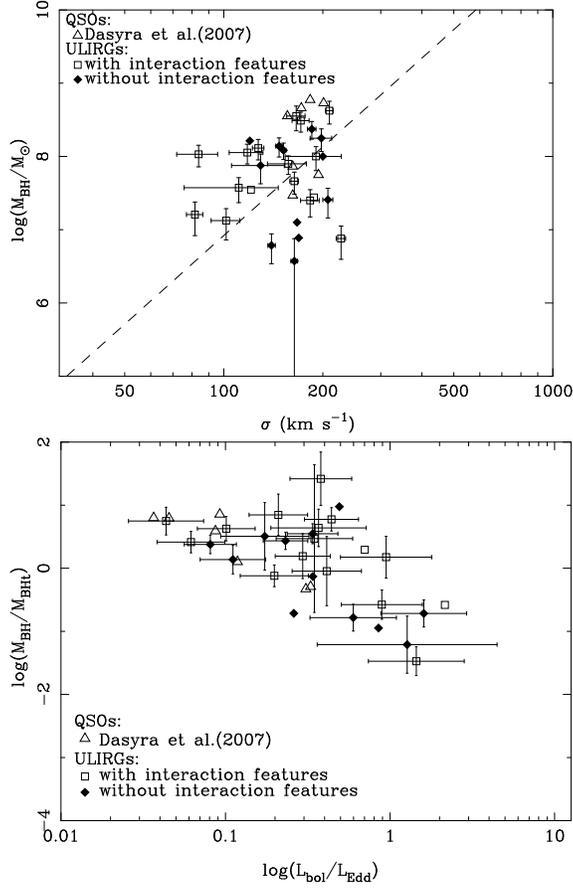

\centering \includegraphics[angle=270,width=75mm]{Mbh_sii_com.ps}\\
\centering \includegraphics[angle=270,width=75mm]{Mbh_Ledd_sii_com.ps}
\caption{$M_{\rm BH}$$-$$\sigma$ and {${\log(M_{\rm BH}/M_{\rm
BHt})}$ vs. ${\log(L_{\rm bol}/L_{\rm Edd})}$
    diagrams} for ULIRGs with or without obvious interaction
  features. {The open squares represent the sources with
    interaction features, the filled rhombuses represent the ULIRGs without
    interaction features, and the open triangles indicate the QSOs with
    measured $\sigma$ in \citet{dtd+07}.}}
\label{mbhcom}
\end{figure}

\begin{figure}
\centering \includegraphics[angle=270,width=75mm]{Mbh_Ledd_NLcore.ps}\\
\centering \includegraphics[angle=270,width=75mm]{Mbh_Ledd_sii.ps}
\caption{${\log(M_{\rm BH}/M_{\rm BHt})}$ vs. ${
    \log(L_{\rm bol}/L_{\rm Edd})}$ relation of Type I ULIRGs. {The
    filled triangles represent the six sources with measured $\sigma$ in
    \citet{dtd06,dtd+06,dtd+07}, the open triangles represent the seven QSOs
    with measured $\sigma$ in \citet{dtd+07}, the filled circles indicate
    the ULIRGs whose parameters are estimated in this work.} Upper
  panel: the M$_{\rm BHt}$ is derived from the $\sigma$ used in
  {the upper panel} of Figure~\ref{mbhoiii}; Lower panel: the
  $M_{\rm BHt}$ is obtained from the $\sigma$ used in {the
    lower }panel of Figure~\ref{mbhoiii}.}
\label{mbhedd}
\end{figure}

\subsection{The $M_{\rm BH}$$-$$\sigma$ relation for Type I ULIRGs}
The $M_{\rm BH}$$-$$\sigma$ relation for ULIRGs was seldom discussed
in the past, mainly due to the poor understanding of BH masses and
the limitation of the sample size. \citet{dtd+06} measured the
velocity dispersions of 54 ULIRGs, and carried out a simulation to
study the reasonableness of using the $M_{\rm BH}$$-$$\sigma$
relation to estimate their BH masses. They concluded that if the
efficiency of gas accretion onto the BH from its surroundings
remains constant with time, this relation can be used. For Type I
ULIRGs, we can estimate their BH masses and velocity dispersions by
use of their optical spectra, thus it is possible to test the
$M_{\rm BH}$$-$$\sigma$ relation for ULIRGs {\it by observation.}
Here we investigate the $M_{\rm BH}$$-$$\sigma$ relation for our
Type I ULIRG sample.

When the stellar velocity dispersions have measured data in the
literature, we adopt them directly, otherwise, the line width of
some narrow emission lines, such as [O$_{\rm III}$]5007, [O$_{\rm
III}$]5007 NL core, [S$_{\rm II}$]6716 or 6731, are used to be the
surrogates for $\sigma$, which were usually adopted in the studies
on different types of AGNs
\citep[e.g.][]{nels00,boro03,gm04,ssgb07,kx07}. For six sources,
F01572+0009, F12540+5708, F15462$-$0450, F21219$-$1757,
F13451+1232(w), and PG 0050+124, their velocity dispersions have
been measured in \citet{dtd06,dtd+06,dtd+07}, and are directly
adopted here. { BH mass uncertainty is estimated from the
  uncertainty of the formula Equation~(15) and the errors of the ${
    L_{\lambda}({\rm 5100{\AA}})}$ and FWHM$_{\rm H_{\beta}}$. The
  uncertainty of $\sigma$ is estimated from the fitting error of FWHM
  of emission lines, and the instrumental resolution for SDSS
  spectra is about 70 km~s$^{-1}$. } To make a comparison, we also use
{the value of} seven QSOs from \citet{dtd+07}, who measured the
$\sigma$ for 11 QSOs. {Among these 11 QSOs,} PG0050+124 is
identified as a Type I ULIRG and PG1404+226 as a NL Seyfert 1 galaxy
in \citet{hxm+05}. PG1426+015 is an interacting system. The BH mass
of LBQS0307$-$0101 is unavailable. Thus only seven of 11 QSOs are
used. {The instrumental resolution for their measured
  velocity dispersion is about 30 km~s$^{-1}$}.

We show the $M_{\rm BH}$$-$$\sigma$ relation of our sample by using
these surrogates of $\sigma$ in Figure~\ref{mbhoiii}, where
$\sigma={\rm
  FWHM_{lines}/2.35}$.  Because the outflow may influence the profile
of [O$_{\rm III}$], there are some problems when using the FWHM of
the [O$_{\rm III}$]5007 profile as the surrogate of $\sigma$. {Thus
we do not adopt
  the FWHM of the [O$_{\rm III}$]5007 profile as the surrogate of $\sigma$.  In the
  upper panel,} the FWHM of the [O$_{\rm III}$] NL core is used {to
  estimate} $\sigma$. \citet{kx07} used [S$_{\rm II}$] as the surrogate of
$\sigma$ and found that NL Seyfert 1s do follow the $
  M_{\rm BH}$$-$$\sigma$ relation. {In the middle panel}, we use the
FWHM of [S$_{\rm II}$] for $\sigma$ of some ULIRGs. Some ULIRGs do
not follow the $M_{\rm BH}$$-$$\sigma$ relation well. The larger
deviations from this relation are, i.e., F11394+0108
  and F17234+6228, whose FWHMs of [S$_{\rm II}$] are broader than those of
[O$_{\rm III}$] profile. The FWHM of [O$_{\rm III}$] NL core is used
to estimate $\sigma$ for these ULIRGs and the result is shown {in
  the lower panel} of Figure~\ref{mbhoiii}. {Considering the
  error bars, we found that most ULIRGs follow the $
    M_{\rm BH}$$-$$\sigma$ relation, but several} sources have
relatively larger deviations compared with those seven QSOs. This
can be seen even from those six ULIRGs with measured velocity
dispersions. Therefore most Type I ULIRGs follow the $M_{\rm
BH}$$-$$\sigma$ relation and {probably are at a late evolutionary
stage, though several ULIRGs do not
  follow this relation, which probably hints that AGN phase
  appears in the late evolutionary stage of ULIRGs, and the $
    M_{\rm BH}$$-$$\sigma$ relation of most Type I ULIRGs has already built up}. According to the popular evolutionary picture of ULIRGs, during the
evolution phase of ULIRGs to QSO, their BHs are growing, and the $
  M_{\rm BH}$$-$$\sigma$ relation is building up step by step. The scatter
in the $M_{\rm BH}$$-$$\sigma$ diagram probably reflect the fact
that ULIRGs are in different evolutionary stages and the central BHs
of some ULIRGs are still rapidly growing. Therefore, the interaction
properties of these Type I ULIRGs may be the hint in understanding
this problem.

In our sample, F13451+3932 is an interaction system with two bright
nuclei, BH mass and velocity dispersion of the west nucleus can be
obtained. F10531+5531 and F15529+4545 are interaction systems.
F14394+5332 interacts with a companion galaxy. Some sources show
that tidal tail/plume features interact with a much smaller galaxy,
such as F01572+0009, F09591+2045, F11162+6020, F07548+4227,
F11134+0225, F14026+4341, F14315+2955,
  F14390+6209  F11206+3639, F11394+0108, F13342+3932, F15320+0325,
F15437+4647, F16122+1531, F17234+6228. Some Type I ULIRGs {are
  probably} interacting systems but not certain because of the lower
image resolution and the lack of redshift data of their companions,
such as F11553$-$0259, F10015$-$0018, F14402+0108. Other ULIRGs do
not show obvious interaction features at least from their SDSS
images. We compare those ULIRGs with or without interaction features
in the $M_{\rm BH}$$-$$\sigma$ diagram. The result is shown in
Figure~\ref{mbhcom}. It seems that the ULIRGs with interaction
features have slightly larger deviations, but the difference is not
significant. Even for those ULIRGs with interaction features, some
of them are close to the $M_{\rm BH}$$-$$\sigma$ line, and others
have larger deviations, which implies that the $M_{\rm
BH}$$-$$\sigma$ relation for some Type I ULIRGs is not fully built
up. The sources in our sample do not show large deviations from the
$M_{\rm BH}$$-$$\sigma$ line, probably because the AGN phase appears
in the late stage of mergers, and the interaction between central BH
and its host galaxy already starts before the ULIRGs present broad
emission lines in the optical band. More observations are needed to
verify the interaction properties and merger stages of all these
Type I ULIRGs.

\subsection{Dependence on the Eddington ratio}
We show the ${\log(M_{\rm BH}/M_{\rm BHt})}$ versus ${
  \log(L_{\rm bol}/L_{\rm Edd})}$ diagram in Figure~\ref{mbhedd}, where $M_{\rm
  BHt}$ is the BH mass calculated from the $M_{\rm BH}$$-$$\sigma$
relation of \citet{tgb+02}. Here ${\log(M_{\rm BH}/M_{\rm BHt})}$
represents the deviation of BH mass from the $M_{\rm BH}$$-$$\sigma$
relation. The Eddington ratio ${\log(L_{\rm bol}/L_{\rm Edd})}$
usually measures the accretion rate of BH. { The uncertainty of
${\log(M_{\rm BH}/M_{\rm BHt})}$ is estimated from the uncertainties
of
  $M_{\rm BH}$, the $M_{\rm BH}$$-$$\sigma$ relation of
  \citet{tgb+02}, and $\sigma$. The uncertainty of ${\log(L_{\rm bol}/L_{\rm Edd})}$ is
  estimated from the errors of $M_{\rm BH}$ and $f_{\rm 5100{\rm
        \AA}}$.} An anti-correlation trend appears in this plot even
  using different methods to estimate $\sigma$. For the sources with
  larger deviations from the $M_{\rm BH}$$-$$\sigma$ relation, they tend
  to have larger Eddington ratios and smaller BH masses, implying that
  their BHs grow faster. For the sources that are close to the $M_{\rm BH}$$-$$\sigma$ relation, they have relatively smaller Eddington
  ratios and larger BH masses, and their central BHs grow slowly. These
  results imply that different kinds of ULIRGs probably exist
  (i.e., some are close to the $M_{\rm BH}$$-$$\sigma$ relation and with a relatively smaller Eddington ratio, some with larger deviations to
  the $M_{\rm BH}$$-$$\sigma$ relation and a relatively larger Eddington
  ratio), and they may have evolutionary connection.  One possible
  explanation of this trend is that the merger of gas-rich galaxies
  will first form a non-regular host galaxy with a larger deviation
  from the $M_{\rm BH}$$-$$\sigma$ relation. The interaction between
  central BH and its host galaxy
  will slowly make it close to the $M_{\rm BH}$$-$$\sigma$ relation,
  since the Eddington
  ratio becomes smaller, the central BH becomes larger and the stars
  around the central region form a more regular spheroid step by
  step. {In Figure~\ref{mbhcom}, we also plot the ${\log(M_{\rm BH}/M_{\rm BHt})}$ versus ${\log(L_{\rm bol}/L_{\rm Edd})}$
    relation for ULIRGs with or without obvious interaction features.}

To estimate $M_{\rm BH}$ and $L_{\rm bol}$/$L_{\rm Edd}$, the
continuum flux at 5100${\rm \AA}$ was used. One may worry that the
trend appeared in the ${\log(M_{\rm BH}/M_{\rm BHt})}$ versus
${\log(L_{\rm bol}/L_{\rm Edd})}$ plot is due to the common
dependence of these two quantities on the continuum flux at
5100${\rm \AA}$. Because ${\log(M_{\rm BH}/M_{\rm BHt})}$ is
proportional to ${\log(L_{{\rm 5100{\rm \AA}}}^{0.61})}$ (we used
the $R-L$ relation of Kaspi et al. 2000), and ${\log(L_{\rm
bol}/L_{\rm Edd})}$ is proportional to ${\log(L_{{\rm 5100{\rm \rm
\AA}}}^{0.39})}$. If the relation depends on the methods of
estimating $M_{\rm BH}$ and $L_{\rm bol}$/$L_{\rm Edd}$,
${\log(M_{\rm BH}/M_{\rm BHt})}$ should increase as ${\rm
\log(L_{bol}/L_{Edd})}$ increases. However, this is not the case in
our plot. Therefore, we believe that the anti-correlation trend
between ${\log(M_{\rm BH}/M_{\rm BHt})}$ and ${\log(L_{\rm
bol}/L_{\rm Edd})}$ is real.

\begin{figure}
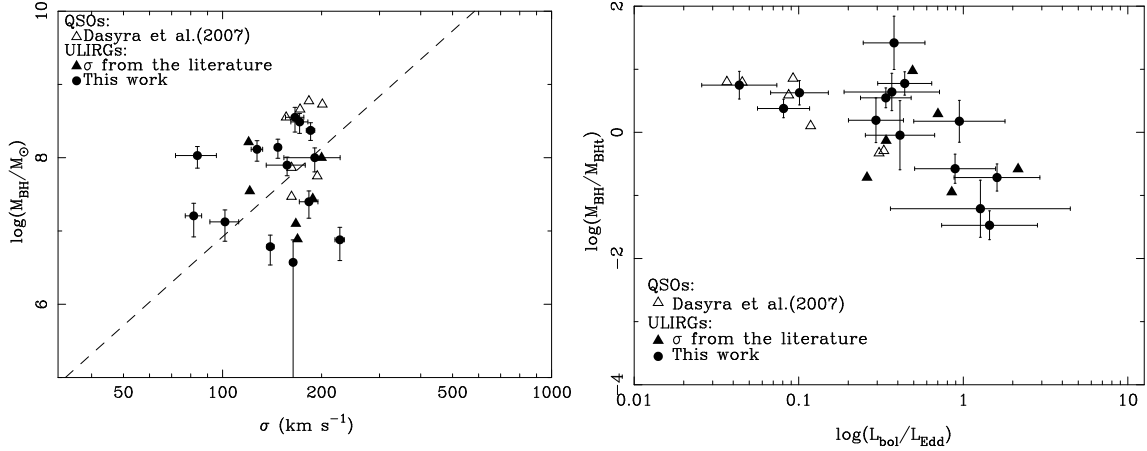

\centering \includegraphics[angle=270,width=75mm]{Mbh_sii_reli.ps}
\centering \includegraphics[angle=270,width=75mm]{Mbh_Ledd_sii_reli.ps}\\

\caption{Two examples about the same results in
Figures~\ref{mbhoiii} and~\ref{mbhedd}, but only the data have been
confirmed in previous
  works were used.}
\label{mbhreli}
\end{figure}

Another important issue is the identification reliability of our
Type I ULIRG sample. For {41} of {62} Type I ULIRGs, their
counterparts and redshifts listed in the NED are consistent with our
results. For other {21} ULIRGs, no counterparts and redshifts are
available in the NED. As discussed in Section 2.3 and also see
Figure~\ref{reli}, {18} of these {21} ULIRGs have their reliability
 greater than {93\%}. Because the large position error of \emph{IRAS}
measurement, and the incompleteness of SDSS spectra, several ULIRGs
identified in this work may be problematic. This will not affect the
statistical results about the NL ULIRGs. But for Type I ULIRGs, we
use {41} ULIRGs which are listed in the NED and six ULIRGs from the
literature \citep{dtd06,dtd+06,dtd+07} to re-do the same work, and
found that our results do not change, as shown in
Figure~\ref{mbhreli}.

\subsection{The relation between ${L_{\rm IR}}$ and ${\lambda L_{\lambda}({5100{\rm \AA}})}$}

\begin{figure}
\centering \includegraphics[angle=270,width=75mm]{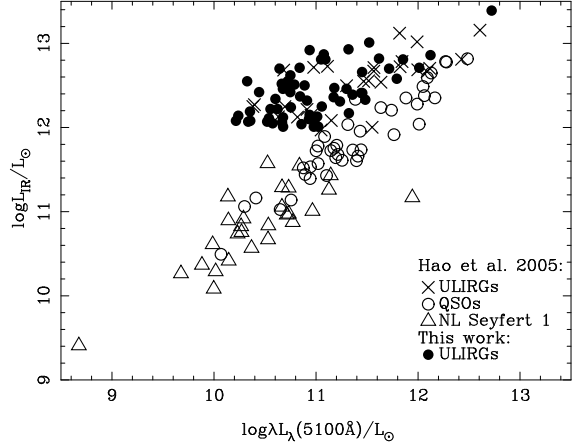}
\caption{$L_{\rm IR}$ vs. ${\lambda L_{\lambda}({\rm 5100{\AA}}})$.
The open circles indicate the PG QSOs, the open triangles represent
the NL Seyfert 1 galaxies, and the crosses denote the Type I ULIRGs.
   These data are taken from \citet{hxm+05}. The filled circles are new
  data of the Type I ULIRGs in our sample.}
\label{lir_lopt}
\end{figure}

The relation between IR and optical continuum luminosities is
important in understanding the origin of the IR emissions in
galaxies. Here we compare the samples of \citet{hxm+05} with our
Type I ULIRGs, because they used the same cosmology as ours,
although their calculation of $L_{\rm IR}$ is not consistent with
ours ({but the difference is very small, about 2\%}).  We show the
$L_{\rm IR}$ versus ${\lambda L_{\lambda}({5100{\rm \AA}})}$ diagram
in Figure~\ref{lir_lopt}. There is a tight correlation between the
$L_{\rm IR}$ and ${\lambda L_{\lambda}({5100{\rm \AA}})}$ for PG
QSOs and NL Seyfert 1 galaxies. The distribution of our {62 Type I
ULIRGs is consistent with that of} \citet{hxm+05}, and most Type I
ULIRGs are above the trend which was defined by the PG QSOs and
narrow line Seyfert 1 galaxies. If this correlation can be explained
as both the infrared and optical continuum emissions of PG QSOs and
NL Seyfert 1 are mainly from the AGN, the deviation to this
correlation may imply an additional contribution to the $L_{\rm IR}$
besides the AGN in Type I ULIRGs, i.e., the contribution from
starbursts.

\section{Discussions}
The intrinsic extinction of ULIRGs is significant. \citet{kan+06}
estimated the optical extinction of three Type I ULIRGs using the
X-ray data, and concluded that the optical extinction $A_{V}$ is
probably less than 1.  While the typical optical extinction
estimated from the Balmer decrement method for the NL ULIRGs is
about 3 \citep[see the $E(B-V)$ values provided in ][under the
assumption of $R$ = $A_{V}$/$E(B-V$) = 3.1]{vks99}. This result
proves that the optical extinction of Type I ULIRGs is typically
smaller than that of NL ULIRGs. We compared the optical extinction
obtained from the X$-$ray method and the Balmer decrement method for
these three Type I ULIRGs. For F11598$-$0112, the optical
extinctions derived by these two methods are consistent with each
other, both being about 0.03. For F01572+0009, these two methods
give smaller values of the optical extinction. For the remaining
one, F11119+3257, the extinctions derived by these two methods have
a large difference. The extinction effect in different evolutionary
stages may have a large difference.

{Recently,} \citet{dww+08} carried out a study on the broad-line
Balmer decrement in Seyfert 1 galaxies and QSOs, and concluded that
the Balmer decrement is a good indicator for dust extinction. We use
the Balmer decrement method to test the influence of extinction on
our results. After such a correction, the mean BH mass of our Type I
ULIRGs sample becomes ${\log(M_{\rm BH}/M_{\odot})\approx 8.1}$,
which is larger than ${\log(M_{\rm BH}/M_{\odot})\approx 7.8}$
obtained without this correction, and is still typically smaller
than that of PG QSOs (${\log(M_{\rm BH}/M_{\odot})\approx 8.4}$).
The anti-correlation trend appeared in the ${\log(M_{\rm BH}/M_{\rm
BHt})}$ versus ${\log(L_{\rm bol}/L_{\rm Edd})}$ plot still exists.
Therefore, we believe that the optical extinction does not affect
our main results significantly.


In Figure~\ref{mbhoiii} and Figure~\ref{mbhedd}, we note that the
sources which are close to the $M_{\rm BH}$$-$$\sigma$ relation tend
to have relatively smaller Eddington ratios and larger BH masses.
The distributions of their BH masses and Eddington ratios are
similar with those of PG QSOs (see the upper panel in
Figure~\ref{mbhedd}). For the ULIRGs with larger deviations from the
$M_{\rm BH}$$-$$\sigma$ line, they tend to have larger Eddington
ratios and smaller BH masses. An evolutionary connection may exist
between them. When the BH activity in ULIRGs is active enough, the
central BH can blow away a large fraction of the surrounding gases,
and the ULIRGs appear as Type I ULIRGs, their central BHs are still
experiencing fast growth (i.e., with a large value of Eddington
ratio), and they have relatively larger deviations from the typical
$M_{\rm BH}$$-$$\sigma$ relation.  As the co-evolution of the
central BH and its host galaxy, the host galaxy will form a spheroid
gradually, and the growth of central BH becomes slower. At this
time, the ULIRGs probably appear like optically bright QSOs. Thus
the evolution of Type I ULIRGs {is} probably followed by the
building up of the $M_{\rm BH}$$-$$\sigma$ relation and evolving
towards the QSO phase.

\citet{cxw+08} compared the mid-infrared spectroscopic properties of
19 local Type I ULIRGs (named IR QSOs in their work) with that of
QSOs and ULIRGs. They concluded that the MIR spectra slopes, the
polycyclic aromatic hydrocarbon (PAH) emission strengths, and
[Ne$_{\rm II}$] 12.81 $\mu$m luminosities of Type I ULIRGs differ
from those of PG QSOs but are comparable with ULIRGs. Their results
support that Type I ULIRGs are at a transitional stage from ULIRGs
to classical QSOs, which are consistent with ours. Therefore, Type I
ULIRGs are important for the understanding of co-evolution {of the}
central BH and its host galaxy.

\section{Conclusions}

We identified {308} ULIRGs from the SDSS DR6 spectroscopy catalog
and the \emph{IRAS} Faint Source Catalog. This is, to date, the
largest ULIRG sample with optical spectra in the local universe.
About {56\%} of them show obvious interaction features, and this
percentage increases for objects with smaller redshifts, probably
due to the selection effect. After excluding the ULIRGs without
obvious emission lines {(with S/N $<$ 3 or suffer serious
absorptions around the emission lines)}, we obtained 147 NL ULIRGs
(with only narrow emission lines in their spectra) and {62} Type I
ULIRGs (with broad emission lines) in this sample. At least {49\%}
of these {209} ULIRGs contain AGNs, and this percentage increases
with L$_{\rm IR}$. The ULIRGs near the radio$-$FIR relation mainly
have the starburst related emissions, and the radio excess in some
objects is due to the AGN related radio emissions.

In combination with previous data, a large sample of {90} Type I
ULIRGs is used to study their BH masses and BH$-$host galaxy
relation. We found that their BH masses are systematically smaller
than those of PG QSOs, even though the optical extinction effects
were corrected through the Balmer decrement method. Their Eddington
ratios are systematically larger than those of PG QSOs. Most Type I
ULIRGs in our sample follow the $M_{\rm BH}$$-$$\sigma$ relation,
but some do not, even the FWHM of [O$_{\rm III}$] narrow line core
or [S$_{\rm II}$] was used as the surrogate of velocity dispersion,
which implies that some Type I ULIRGs are in the early stage of
galaxy evolution and their $M_{\rm BH}$$-$$\sigma$ relation is still
building up.

We found an anti-correlation trend between ${\log(M_{\rm BH}/M_{\rm
BHt})}$ and ${\log(L_{\rm bol}/L_{\rm Edd})}$, where $M_{\rm BHt}$
is the BH mass derived from the $M_{\rm BH}$$-$$\sigma$ relation of
\citet{tgb+02}. The Type I ULIRGs with larger deviations from the
$M_{\rm BH}$$-$$\sigma$ relation tend to have larger Eddington
ratios and smaller BH masses, and the ULIRGs which are close to the
$M_{\rm BH}$$-$$\sigma$ relation tend to have smaller Eddington
ratios and larger BH masses. This anti-correlation trend implies
that the evolution of Type I ULIRGs {is} probably followed by the
building up of the $M_{\rm BH}$$-$$\sigma$ relation and evolving
towards the QSO phase. Different types of ULIRGs are probably at
different evolution stages.

\begin{acknowledgements}

  {We thank the referee for instructive comments}, and Xiaoyang
  Xia, Fukun Liu, Hong Wu, Minzhi Kong, Ran Wang, Zhonglue Wen, Hui
  Shi, and Zhaoyu Li for their helpful suggestions and discussions. This
  work is supported by the National Natural Science Foundation of
  China (10525313, 10773016,10821061 and 10833003), and the National
  Key Basic Research Science Foundation of China (2007CB815403,
  2007CB815405). Funding for the SDSS has been provided by the Alfred
  P. Sloan Foundation, the Participating Institutions, the National
  Science Foundation, the Department of Energy, the National
  Aeronautics and Space Administration, the Japanese Monbukagakusho,
  the Max Planck Society, and the Higher Education Funding Council for
  England. The SDSS Web site is http://www.sdss.org. The SDSS is
  managed by the Astrophysical Research Consortium for the
  Participating Institutions. The Participating Institutions are the
  American Museum of Natural History, Astrophysical Institute
  Potsdam, University of Basel, University of Cambridge, CaseWestern
  Reserve University, University of Chicago, Drexel University,
  Fermilab, the Institute for Advanced Study, the Japan Participation
  Group, Johns Hopkins University, the Joint Institute for Nuclear
  Astrophysics, the Kavli Institute for Particle Astrophysics and
  Cosmology, the Korean Scientist Group, the Chinese Academy of
  Sciences (LAMOST), Los Alamos National Laboratory, the Max Planck
  Institute for Astronomy (MPIA), the Max Planck Institute for
  Astrophysics (MPA), NewMexico State University, Ohio State
  University, University of Pittsburgh, University of Portsmouth,
  Princeton University, the United States Naval Observatory, and the
  University of Washington.

\end{acknowledgements}

\bibliographystyle{apj} \bibliography{rev.bib}

\begin{thebibliography}{74}
\expandafter\ifx\csname natexlab\endcsname\relax\def\natexlab#1{#1}\fi

\bibitem[{{Adelman-McCarthy} {et~al.}(2008){Adelman-McCarthy}, {Ag{\"u}eros},
  {Allam}, {Allende Prieto}, {Anderson}, {Anderson}, {Annis}, {Bahcall},
  {Bailer-Jones}, {Baldry}, {Barentine}, {Bassett}, {Becker}, {Beers}, {Bell},
  {Berlind}, {Bernardi}, {Blanton}, {Bochanski}, {Boroski}, {Brinchmann},
  {Brinkmann}, {Brunner}, {Budav{\'a}ri}, {Carliles}, {Carr}, {Castander},
  {Cinabro}, {Cool}, {Covey}, {Csabai}, {Cunha}, {Davenport}, {Dilday}, {Doi},
  {Eisenstein}, {Evans}, {Fan}, {Finkbeiner}, {Friedman}, {Frieman},
  {Fukugita}, {G{\"a}nsicke}, {Gates}, {Gillespie}, {Glazebrook}, {Gray},
  {Grebel}, {Gunn}, {Gurbani}, {Hall}, {Harding}, {Harvanek}, {Hawley},
  {Hayes}, {Heckman}, {Hendry}, {Hindsley}, {Hirata}, {Hogan}, {Hogg}, {Hyde},
  {Ichikawa}, {Ivezi{\'c}}, {Jester}, {Johnson}, {Jorgensen}, {Juri{\'c}},
  {Kent}, {Kessler}, {Kleinman}, {Knapp}, {Kron}, {Krzesinski}, {Kuropatkin},
  {Lamb}, {Lampeitl}, {Lebedeva}, {Lee}, {Leger}, {L{\'e}pine}, {Lima}, {Lin},
  {Long}, {Loomis}, {Loveday}, {Lupton}, {Malanushenko}, {Malanushenko},
  {Mandelbaum}, {Margon}, {Marriner}, {Mart{\'{\i}}nez-Delgado}, {Matsubara},
  {McGehee}, {McKay}, {Meiksin}, {Morrison}, {Munn}, {Nakajima}, {Neilsen},
  {Newberg}, {Nichol}, {Nicinski}, {Nieto-Santisteban}, {Nitta}, {Okamura},
  {Owen}, {Oyaizu}, {Padmanabhan}, {Pan}, {Park}, {Peoples}, {Pier}, {Pope},
  {Purger}, {Raddick}, {Re Fiorentin}, {Richards}, {Richmond}, {Riess}, {Rix},
  {Rockosi}, {Sako}, {Schlegel}, {Schneider}, {Schreiber}, {Schwope}, {Seljak},
  {Sesar}, {Sheldon}, {Shimasaku}, {Sivarani}, {Smith}, {Snedden}, {Steinmetz},
  {Strauss}, {SubbaRao}, {Suto}, {Szalay}, {Szapudi}, {Szkody}, {Tegmark},
  {Thakar}, {Tremonti}, {Tucker}, {Uomoto}, {Vanden Berk}, {Vandenberg},
  {Vidrih}, {Vogeley}, {Voges}, {Vogt}, {Wadadekar}, {Weinberg}, {West},
  {White}, {Wilhite}, {Yanny}, {Yocum}, {York}, {Zehavi}, \& {Zucker}}]{adel08}
{Adelman-McCarthy}, J.~K. et al. 2008, \apjs, 175, 297

\bibitem[{{Alexander} {et~al.}(2008){Alexander}, {Brandt}, {Smail}, {Swinbank},
  {Bauer}, {Blain}, {Chapman}, {Coppin}, {Ivison}, \&
  {Men{\'e}ndez-Delmestre}}]{abs+98}
{Alexander}, D.~M., {Brandt}, W.~N., {Smail}, I., {Swinbank}, A.~M., {Bauer},
  F.~E., {Blain}, A.~W., {Chapman}, S.~C., {Coppin}, K.~E.~K., {Ivison}, R.~J.,
  \& {Men{\'e}ndez-Delmestre}, K. 2008, \aj, 135, 1968

\bibitem[{{Baldwin} {et~al.}(1981){Baldwin}, {Phillips}, \&
  {Terlevich}}]{bpt81}
{Baldwin}, J.~A., {Phillips}, M.~M., \& {Terlevich}, R. 1981, \pasp, 93, 5

\bibitem[{{Becker} {et~al.}(1995){Becker}, {White}, \& {Helfand}}]{bwh95}
{Becker}, R.~H., {White}, R.~L., \& {Helfand}, D.~J. 1995, \apj, 450, 559

\bibitem[{{Best} {et~al.}(2005){Best}, {Kauffmann}, {Heckman}, \&
  {Ivezi{\'c}}}]{bkhi05}
{Best}, P.~N., {Kauffmann}, G., {Heckman}, T.~M., \& {Ivezi{\'c}}, {\v Z}.
  2005, \mnras, 362, 9

\bibitem[{{Boroson}(2003)}]{boro03}
{Boroson}, T.~A. 2003, \apj, 585, 647

\bibitem[{{Boroson} \& {Green}(1992)}]{bg92}
{Boroson}, T.~A. \& {Green}, R.~F. 1992, \apjs, 80, 109

\bibitem[{{Borys} {et~al.}(2005){Borys}, {Smail}, {Chapman}, {Blain},
  {Alexander}, \& {Ivison}}]{bsc+05}
{Borys}, C., {Smail}, I., {Chapman}, S.~C., {Blain}, A.~W., {Alexander}, D.~M.,
  \& {Ivison}, R.~J. 2005, \apj, 635, 853

\bibitem[{{Calzetti} {et~al.}(2000){Calzetti}, {Armus}, {Bohlin}, {Kinney},
  {Koornneef}, \& {Storchi-Bergmann}}]{cab+00}
{Calzetti}, D., {Armus}, L., {Bohlin}, R.~C., {Kinney}, A.~L., {Koornneef}, J.,
  \& {Storchi-Bergmann}, T. 2000, \apj, 533, 682

\bibitem[{{Canalizo} \& {Stockton}(2001)}]{cs01}
{Canalizo}, G. \& {Stockton}, A. 2001, \apj, 555, 719

\bibitem[{{Cao} {et~al.}(2006){Cao}, {Wu}, {Wang}, {Hao}, {Deng}, {Xia}, \&
  {Zou}}]{cww+06}
{Cao}, C., {Wu}, H., {Wang}, J.-L., {Hao}, C.-N., {Deng}, Z.-G., {Xia}, X.-Y.,
  \& {Zou}, Z.-L. 2006, Chinese Journal of Astronomy and Astrophysics, 6, 197

\bibitem[{{Cao} {et~al.}(2008){Cao}, {Xia}, {Wu}, {Mao}, {Hao}, \&
  {Deng}}]{cxw+08}
{Cao}, C., {Xia}, X.~Y., {Wu}, H., {Mao}, S., {Hao}, C.~N., \& {Deng}, Z.~G.
  2008, \mnras, 390, 336

\bibitem[{{Cardelli} {et~al.}(1989){Cardelli}, {Clayton}, \& {Mathis}}]{ccm89}
{Cardelli}, J.~A., {Clayton}, G.~C., \& {Mathis}, J.~S. 1989, \apj, 345, 245

\bibitem[{{Clements} {et~al.}(1996){Clements}, {Sutherland}, {McMahon}, \&
  {Saunders}}]{csm+96}
{Clements}, D.~L., {Sutherland}, W.~J., {McMahon}, R.~G., \& {Saunders}, W.
  1996, \mnras, 279, 477

\bibitem[{{Colless} {et~al.}(2001){Colless}, {Dalton}, {Maddox}, {Sutherland},
  {Norberg}, {Cole}, {Bland-Hawthorn}, {Bridges}, {Cannon}, {Collins}, {Couch},
  {Cross}, {Deeley}, {De Propris}, {Driver}, {Efstathiou}, {Ellis}, {Frenk},
  {Glazebrook}, {Jackson}, {Lahav}, {Lewis}, {Lumsden}, {Madgwick}, {Peacock},
  {Peterson}, {Price}, {Seaborne}, \& {Taylor}}]{cdm+01}
{Colless}, M., {Dalton}, G., {Maddox}, S., {Sutherland}, W., {Norberg}, P.,
  {Cole}, S., {Bland-Hawthorn}, J., {Bridges}, T., {Cannon}, R., {Collins}, C.,
  {Couch}, W., {Cross}, N., {Deeley}, K., {De Propris}, R., {Driver}, S.~P.,
  {Efstathiou}, G., {Ellis}, R.~S., {Frenk}, C.~S., {Glazebrook}, K.,
  {Jackson}, C., {Lahav}, O., {Lewis}, I., {Lumsden}, S., {Madgwick}, D.,
  {Peacock}, J.~A., {Peterson}, B.~A., {Price}, I., {Seaborne}, M., \&
  {Taylor}, K. 2001, \mnras, 328, 1039

\bibitem[{{Condon} {et~al.}(1991{\natexlab{a}}){Condon}, {Anderson}, \&
  {Helou}}]{cah91}
{Condon}, J.~J., {Anderson}, M.~L., \& {Helou}, G. 1991{\natexlab{a}}, \apj,
  376, 95

\bibitem[{{Condon} {et~al.}(1998){Condon}, {Cotton}, {Greisen}, {Yin},
  {Perley}, {Taylor}, \& {Broderick}}]{ccg+98}
{Condon}, J.~J., {Cotton}, W.~D., {Greisen}, E.~W., {Yin}, Q.~F., {Perley},
  R.~A., {Taylor}, G.~B., \& {Broderick}, J.~J. 1998, \aj, 115, 1693

\bibitem[{{Condon} {et~al.}(1991{\natexlab{b}}){Condon}, {Huang}, {Yin}, \&
  {Thuan}}]{chyt91}
{Condon}, J.~J., {Huang}, Z.-P., {Yin}, Q.~F., \& {Thuan}, T.~X.
  1991{\natexlab{b}}, \apj, 378, 65

\bibitem[{{Cui} {et~al.}(2001){Cui}, {Xia}, {Deng}, {Mao}, \& {Zou}}]{cxd+01}
{Cui}, J., {Xia}, X.-Y., {Deng}, Z.-G., {Mao}, S., \& {Zou}, Z.-L. 2001, \aj,
  122, 63

\bibitem[{{Dasyra} {et~al.}(2006{\natexlab{a}}){Dasyra}, {Tacconi}, {Davies},
  {Genzel}, {Lutz}, {Naab}, {Burkert}, {Veilleux}, \& {Sanders}}]{dtd06}
{Dasyra}, K.~M., {Tacconi}, L.~J., {Davies}, R.~I., {Genzel}, R., {Lutz}, D.,
  {Naab}, T., {Burkert}, A., {Veilleux}, S., \& {Sanders}, D.~B.
  2006{\natexlab{a}}, \apj, 638, 745

\bibitem[{{Dasyra} {et~al.}(2007){Dasyra}, {Tacconi}, {Davies}, {Genzel},
  {Lutz}, {Peterson}, {Veilleux}, {Baker}, {Schweitzer}, \& {Sturm}}]{dtd+07}
{Dasyra}, K.~M., {Tacconi}, L.~J., {Davies}, R.~I., {Genzel}, R., {Lutz}, D.,
  {Peterson}, B.~M., {Veilleux}, S., {Baker}, A.~J., {Schweitzer}, M., \&
  {Sturm}, E. 2007, \apj, 657, 102

\bibitem[{{Dasyra} {et~al.}(2006{\natexlab{b}}){Dasyra}, {Tacconi}, {Davies},
  {Naab}, {Genzel}, {Lutz}, {Sturm}, {Baker}, {Veilleux}, {Sanders}, \&
  {Burkert}}]{dtd+06}
{Dasyra}, K.~M., {Tacconi}, L.~J., {Davies}, R.~I., {Naab}, T., {Genzel}, R.,
  {Lutz}, D., {Sturm}, E., {Baker}, A.~J., {Veilleux}, S., {Sanders}, D.~B., \&
  {Burkert}, A. 2006{\natexlab{b}}, \apj, 651, 835

\bibitem[{{Dong} {et~al.}(2008){Dong}, {Wang}, {Wang}, {Yuan}, {Zhou}, {Dai},
  \& {Zhang}}]{dww+08}
{Dong}, X., {Wang}, T., {Wang}, J., {Yuan}, W., {Zhou}, H., {Dai}, H., \&
  {Zhang}, K. 2008, \mnras, 383, 581

\bibitem[{{Farrah} {et~al.}(2001){Farrah}, {Rowan-Robinson}, {Oliver},
  {Serjeant}, {Borne}, {Lawrence}, {Lucas}, {Bushouse}, \& {Colina}}]{fro+01}
{Farrah}, D., {Rowan-Robinson}, M., {Oliver}, S., {Serjeant}, S., {Borne}, K.,
  {Lawrence}, A., {Lucas}, R.~A., {Bushouse}, H., \& {Colina}, L. 2001, \mnras,
  326, 1333

\bibitem[{{Gao} \& {Solomon}(2004)}]{gs04}
{Gao}, Y. \& {Solomon}, P.~M. 2004, \apj, 606, 271

\bibitem[{{Goto}(2005)}]{goto05}
{Goto}, T. 2005, \mnras, 360, 322

\bibitem[{{Grupe} \& {Mathur}(2004)}]{gm04}
{Grupe}, D. \& {Mathur}, S. 2004, \apjl, 606, L41

\bibitem[{{Hao} {et~al.}(2005){Hao}, {Xia}, {Mao}, {Wu}, \& {Deng}}]{hxm+05}
{Hao}, C.~N., {Xia}, X.~Y., {Mao}, S., {Wu}, H., \& {Deng}, Z.~G. 2005, \apj,
  625, 78

\bibitem[{{Helou} {et~al.}(1988){Helou}, {Khan}, {Malek}, \&
  {Boehmer}}]{hkmb88}
{Helou}, G., {Khan}, I.~R., {Malek}, L., \& {Boehmer}, L. 1988, \apjs, 68, 151

\bibitem[{{Helou} {et~al.}(1985){Helou}, {Soifer}, \& {Rowan-Robinson}}]{hsr85}
{Helou}, G., {Soifer}, B.~T., \& {Rowan-Robinson}, M. 1985, \apjl, 298, L7

\bibitem[{{Hwang} {et~al.}(2007){Hwang}, {Serjeant}, {Lee}, {Lee}, \&
  {White}}]{hsl+07}
{Hwang}, H.~S., {Serjeant}, S., {Lee}, M.~G., {Lee}, K.~H., \& {White}, G.~J.
  2007, \mnras, 375, 115

\bibitem[{{Jones} {et~al.}(2004){Jones}, {Saunders}, {Colless}, {Read},
  {Parker}, {Watson}, {Campbell}, {Burkey}, {Mauch}, {Moore}, {Hartley},
  {Cass}, {James}, {Russell}, {Fiegert}, {Dawe}, {Huchra}, {Jarrett}, {Lahav},
  {Lucey}, {Mamon}, {Proust}, {Sadler}, \& {Wakamatsu}}]{jsc+04}
{Jones}, D.~H., {Saunders}, W., {Colless}, M., {Read}, M.~A., {Parker}, Q.~A.,
  {Watson}, F.~G., {Campbell}, L.~A., {Burkey}, D., {Mauch}, T., {Moore}, L.,
  {Hartley}, M., {Cass}, P., {James}, D., {Russell}, K., {Fiegert}, K., {Dawe},
  J., {Huchra}, J., {Jarrett}, T., {Lahav}, O., {Lucey}, J., {Mamon}, G.~A.,
  {Proust}, D., {Sadler}, E.~M., \& {Wakamatsu}, K.-i. 2004, \mnras, 355, 747

\bibitem[{{Kaspi} {et~al.}(2000){Kaspi}, {Smith}, {Netzer}, {Maoz}, {Jannuzi},
  \& {Giveon}}]{ksn+00}
{Kaspi}, S., {Smith}, P.~S., {Netzer}, H., {Maoz}, D., {Jannuzi}, B.~T., \&
  {Giveon}, U. 2000, \apj, 533, 631

\bibitem[{{Kauffmann} {et~al.}(2003){Kauffmann}, {Heckman}, {Tremonti},
  {Brinchmann}, {Charlot}, {White}, {Ridgway}, {Brinkmann}, {Fukugita}, {Hall},
  {Ivezi{\'c}}, {Richards}, \& {Schneider}}]{kht+03}
{Kauffmann}, G., {Heckman}, T.~M., {Tremonti}, C., {Brinchmann}, J., {Charlot},
  S., {White}, S.~D.~M., {Ridgway}, S.~E., {Brinkmann}, J., {Fukugita}, M.,
  {Hall}, P.~B., {Ivezi{\'c}}, {\v Z}., {Richards}, G.~T., \& {Schneider},
  D.~P. 2003, \mnras, 346, 1055

\bibitem[{{Kawakatu} {et~al.}(2006){Kawakatu}, {Anabuki}, {Nagao}, {Umemura},
  \& {Nakagawa}}]{kan+06}
{Kawakatu}, N., {Anabuki}, N., {Nagao}, T., {Umemura}, M., \& {Nakagawa}, T.
  2006, \apj, 637, 104

\bibitem[{{Kewley} {et~al.}(2006){Kewley}, {Groves}, {Kauffmann}, \&
  {Heckman}}]{kgkh06}
{Kewley}, L.~J., {Groves}, B., {Kauffmann}, G., \& {Heckman}, T. 2006, \mnras,
  372, 961

\bibitem[{{Kim} \& {Sanders}(1998)}]{ks98}
{Kim}, D.-C. \& {Sanders}, D.~B. 1998, \apjs, 119, 41

\bibitem[{{Kim} {et~al.}(1998){Kim}, {Veilleux}, \& {Sanders}}]{kvs98}
{Kim}, D.-C., {Veilleux}, S., \& {Sanders}, D.~B. 1998, \apj, 508, 627

\bibitem[{{Komossa} \& {Xu}(2007)}]{kx07}
{Komossa}, S. \& {Xu}, D. 2007, \apjl, 667, L33

\bibitem[{{Lonsdale} {et~al.}(1998){Lonsdale}, {Conrow}, {Evans}, {Fullmer},
  {Moshir}, {Chester}, {Yentis}, {MacGillivray}, {Wolstencroft}, \&
  {Egret}}]{lce+98}
{Lonsdale}, C., {Conrow}, T., {Evans}, T., {Fullmer}, L., {Moshir}, M.,
  {Chester}, T., {Yentis}, D., {MacGillivray}, H., {Wolstencroft}, R., \&
  {Egret}, D. 1998, in IAU Symposium, Vol. 179, New Horizons from
  Multi-Wavelength Sky Surveys, ed. B.~J. {McLean}, D.~A. {Golombek}, J.~J.~E.
  {Hayes}, \& H.~E. {Payne}, 450--+

\bibitem[{{Lonsdale} {et~al.}(2006){Lonsdale}, {Farrah}, \& {Smith}}]{lfs06}
{Lonsdale}, C.~J., {Farrah}, D., \& {Smith}, H.~E. 2006, {Ultraluminous
  Infrared Galaxies} (Astrophysics Update 2), 285--+

\bibitem[{{Masci} {et~al.}(2001){Masci}, {Condon}, {Barlow}, {Lonsdale}, {Xu},
  {Shupe}, {Pevunova}, {Fang}, \& {Cutri}}]{mcb+01}
{Masci}, F.~J., {Condon}, J.~J., {Barlow}, T.~A., {Lonsdale}, C.~J., {Xu}, C.,
  {Shupe}, D.~L., {Pevunova}, O., {Fang}, F., \& {Cutri}, R. 2001, \pasp, 113,
  10

\bibitem[{{McLure} \& {Jarvis}(2002)}]{mj02}
{McLure}, R.~J. \& {Jarvis}, M.~J. 2002, \mnras, 337, 109

\bibitem[{{Mirabel} \& {Sanders}(1988)}]{ms88}
{Mirabel}, I.~F. \& {Sanders}, D.~B. 1988, \apj, 335, 104

\bibitem[{{Moshir} {et~al.}(1992){Moshir}, {Kopman}, \& {Conrow}}]{mkc92}
{Moshir}, M., {Kopman}, G., \& {Conrow}, T.~A.~O. 1992, {IRAS Faint Source
  Survey, Explanatory supplement version 2} (Pasadena: Infrared Processing and
  Analysis Center, California Institute of Technology, 1992, edited by Moshir,
  M.; Kopman, G.; Conrow, T.~a.o.)

\bibitem[{{Murphy} {et~al.}(1996){Murphy}, {Armus}, {Matthews}, {Soifer},
  {Mazzarella}, {Shupe}, {Strauss}, \& {Neugebauer}}]{mam+96}
{Murphy}, Jr., T.~W., {Armus}, L., {Matthews}, K., {Soifer}, B.~T.,
  {Mazzarella}, J.~M., {Shupe}, D.~L., {Strauss}, M.~A., \& {Neugebauer}, G.
  1996, \aj, 111, 1025

\bibitem[{{Nelson}(2000)}]{nels00}
{Nelson}, C.~H. 2000, \apjl, 544, L91

\bibitem[{{Pasquali} {et~al.}(2005){Pasquali}, {Kauffmann}, \&
  {Heckman}}]{pkh05}
{Pasquali}, A., {Kauffmann}, G., \& {Heckman}, T.~M. 2005, \mnras, 361, 1121

\bibitem[{{Peterson} \& {Wandel}(1999)}]{pw99}
{Peterson}, B.~M. \& {Wandel}, A. 1999, \apjl, 521, L95

\bibitem[{{Peterson} \& {Wandel}(2000)}]{pw00}
---. 2000, \apjl, 540, L13

\bibitem[{{Roy} \& {Norris}(1997)}]{rn97}
{Roy}, A.~L. \& {Norris}, R.~P. 1997, \mnras, 289, 824

\bibitem[{{Salviander} {et~al.}(2007){Salviander}, {Shields}, {Gebhardt}, \&
  {Bonning}}]{ssgb07}
{Salviander}, S., {Shields}, G.~A., {Gebhardt}, K., \& {Bonning}, E.~W. 2007,
  \apj, 662, 131

\bibitem[{{Sanders} {et~al.}(2003){Sanders}, {Mazzarella}, {Kim}, {Surace}, \&
  {Soifer}}]{smk+03}
{Sanders}, D.~B., {Mazzarella}, J.~M., {Kim}, D.-C., {Surace}, J.~A., \&
  {Soifer}, B.~T. 2003, \aj, 126, 1607

\bibitem[{{Sanders} \& {Mirabel}(1996)}]{sm96}
{Sanders}, D.~B. \& {Mirabel}, I.~F. 1996, \araa, 34, 749

\bibitem[{{Sanders} {et~al.}(1991){Sanders}, {Scoville}, \& {Soifer}}]{sss91}
{Sanders}, D.~B., {Scoville}, N.~Z., \& {Soifer}, B.~T. 1991, \apj, 370, 158

\bibitem[{{Sanders} {et~al.}(1986){Sanders}, {Scoville}, {Young}, {Soifer},
  {Schloerb}, {Rice}, \& {Danielson}}]{ssy+86}
{Sanders}, D.~B., {Scoville}, N.~Z., {Young}, J.~S., {Soifer}, B.~T.,
  {Schloerb}, F.~P., {Rice}, W.~L., \& {Danielson}, G.~E. 1986, \apjl, 305, L45

\bibitem[{{Soifer} {et~al.}(1987){Soifer}, {Sanders}, {Madore}, {Neugebauer},
  {Danielson}, {Elias}, {Lonsdale}, \& {Rice}}]{ssm+87}
{Soifer}, B.~T., {Sanders}, D.~B., {Madore}, B.~F., {Neugebauer}, G.,
  {Danielson}, G.~E., {Elias}, J.~H., {Lonsdale}, C.~J., \& {Rice}, W.~L. 1987,
  \apj, 320, 238

\bibitem[{{Stanford} {et~al.}(2000){Stanford}, {Stern}, {van Breugel}, \& {De
  Breuck}}]{ssbb00}
{Stanford}, S.~A., {Stern}, D., {van Breugel}, W., \& {De Breuck}, C. 2000,
  \apjs, 131, 185

\bibitem[{{Strauss} {et~al.}(1990){Strauss}, {Davis}, {Yahil}, \&
  {Huchra}}]{sdyh90}
{Strauss}, M.~A., {Davis}, M., {Yahil}, A., \& {Huchra}, J.~P. 1990, \apj, 361,
  49

\bibitem[{{Surace} {et~al.}(2000){Surace}, {Sanders}, \& {Evans}}]{sse00}
{Surace}, J.~A., {Sanders}, D.~B., \& {Evans}, A.~S. 2000, \apj, 529, 170

\bibitem[{{Sutherland} \& {Saunders}(1992)}]{ss92}
{Sutherland}, W. \& {Saunders}, W. 1992, \mnras, 259, 413

\bibitem[{{Tremaine} {et~al.}(2002){Tremaine}, {Gebhardt}, {Bender}, {Bower},
  {Dressler}, {Faber}, {Filippenko}, {Green}, {Grillmair}, {Ho}, {Kormendy},
  {Lauer}, {Magorrian}, {Pinkney}, \& {Richstone}}]{tgb+02}
{Tremaine}, S., {Gebhardt}, K., {Bender}, R., {Bower}, G., {Dressler}, A.,
  {Faber}, S.~M., {Filippenko}, A.~V., {Green}, R., {Grillmair}, C., {Ho},
  L.~C., {Kormendy}, J., {Lauer}, T.~R., {Magorrian}, J., {Pinkney}, J., \&
  {Richstone}, D. 2002, \apj, 574, 740

\bibitem[{{Veilleux} {et~al.}(1999{\natexlab{a}}){Veilleux}, {Kim}, \&
  {Sanders}}]{vks99}
{Veilleux}, S., {Kim}, D.-C., \& {Sanders}, D.~B. 1999{\natexlab{a}}, \apj,
  522, 113

\bibitem[{{Veilleux} {et~al.}(2002){Veilleux}, {Kim}, \& {Sanders}}]{vks02}
---. 2002, \apjs, 143, 315

\bibitem[{{Veilleux} {et~al.}(1995){Veilleux}, {Kim}, {Sanders}, {Mazzarella},
  \& {Soifer}}]{vks+95}
{Veilleux}, S., {Kim}, D.-C., {Sanders}, D.~B., {Mazzarella}, J.~M., \&
  {Soifer}, B.~T. 1995, \apjs, 98, 171

\bibitem[{{Veilleux} {et~al.}(1997){Veilleux}, {Sanders}, \& {Kim}}]{vsk97}
{Veilleux}, S., {Sanders}, D.~B., \& {Kim}, D.-C. 1997, \apj, 484, 92

\bibitem[{{Veilleux} {et~al.}(1999{\natexlab{b}}){Veilleux}, {Sanders}, \&
  {Kim}}]{vsk99}
---. 1999{\natexlab{b}}, \apj, 522, 139

\bibitem[{{White} {et~al.}(1997){White}, {Becker}, {Helfand}, \&
  {Gregg}}]{wbhg97}
{White}, R.~L., {Becker}, R.~H., {Helfand}, D.~J., \& {Gregg}, M.~D. 1997,
  \apj, 475, 479

\bibitem[{{Wu} {et~al.}(1998{\natexlab{a}}){Wu}, {Zou}, {Xia}, \&
  {Deng}}]{wzxd981}
{Wu}, H., {Zou}, Z.~L., {Xia}, X.~Y., \& {Deng}, Z.~G. 1998{\natexlab{a}},
  \aaps, 127, 521

\bibitem[{{Wu} {et~al.}(1998{\natexlab{b}}){Wu}, {Zou}, {Xia}, \&
  {Deng}}]{wzxd98}
---. 1998{\natexlab{b}}, \aaps, 132, 181

\bibitem[{{Yun} {et~al.}(2001){Yun}, {Reddy}, \& {Condon}}]{yrc01}
{Yun}, M.~S., {Reddy}, N.~A., \& {Condon}, J.~J. 2001, \apj, 554, 803

\bibitem[{{Zheng} {et~al.}(2002){Zheng}, {Xia}, {Mao}, {Wu}, \&
  {Deng}}]{zxm+02}
{Zheng}, X.~Z., {Xia}, X.~Y., {Mao}, S., {Wu}, H., \& {Deng}, Z.~G. 2002, \aj,
  124, 18

\bibitem[{{Zheng} {et~al.}(1999){Zheng}, {Wu}, {Mao}, {Xia}, {Deng}, \&
  {Zou}}]{zwm+99}
{Zheng}, Z., {Wu}, H., {Mao}, S., {Xia}, X.-Y., {Deng}, Z.-G., \& {Zou}, Z.-L.
  1999, \aap, 349, 735

\bibitem[{{Zou} {et~al.}(1991){Zou}, {Xia}, {Deng}, \& {Su}}]{zxds91}
{Zou}, Z., {Xia}, X., {Deng}, Z., \& {Su}, H. 1991, \mnras, 252, 593

\end{thebibliography}
\appendix
\section{The parameters of our samples (online version)}



\end{document}